\documentclass[useAMS,usenatbib]{mn2e}
\usepackage{graphicx,psfig,cite,epsf}  
\usepackage{times}
\usepackage{subfigure}
\usepackage{textcomp}
\usepackage{multirow}

\def\apjl{ApJ}
\def\apjl{ApJ}                
\def\apjs{ApJS}               
\def\ao{Appl.~Opt.}           
\def\aap{A\&A}                

\title{Ultra Luminous X-ray Sources: a deeper insight into their spectral evolution}
\author[Fabio Pintore, Luca Zampieri, Anna Wolter, Tomaso Belloni]{Fabio Pintore$^1$, Luca Zampieri$^1$, Anna Wolter$^2$, Tomaso Belloni$^3$\\
$^1$ INAF-Osservatorio Astronomico di Padova, Vicolo dell'Osservatorio 5, I-35122 Padova, Italy \\
$^2$ INAF, Osservatorio Astronomico di Brera, via Brera 28, 20121 Milano, Italy\\
$^3$ INAF, Osservatorio Astronomico di Brera, via E. Bianchi 46, I-23807 Merate, Italy}

\begin{document}

\maketitle

\begin{abstract}

{We select a sample of nearby Ultraluminous X-ray sources with long XMM-Newton observations and analyse all the available XMM-Newton data using both X-ray spectral fitting techniques and hardness-intensity diagrams.
The sample includes IC 342 X-1, NGC 5204 X-1, NGC 5408 X-1, Holmberg IX X-1, Holmberg II X-1, NGC 1313 X-1, NGC 1313 X-2 and NGC 253 X-1.
We found that, although a common reference model can be used to describe the X-ray spectra, the sources show different spectral evolutions, phenomenologically described in terms of variations in the properties of a soft component and a high energy tail. Variations at low energies are accounted for (mostly) by changes in the normalization of the soft component and/or in the column density to the source, while variations in the high energy tail by changes in the parameters of an optically thick corona. This spectral variability is rather well characterized on a colour-colour and hardness-intensity diagram in terms of suitably defined hardness ratios. We suggest the existence of a variability pattern on the hardness-intensity diagram and we interpret it in terms of the switch between a near-Eddington and a super-Eddington accretion regime. The transition between the two regimes seems to be driven mostly by changes in the contribution of the soft component, which can be explained in terms of the increasing importance of wind emission. The analysis is complemented by an investigation of the short-term time variability of all the sources. In general, no clear correlation between the spectral and temporal properties is found.}
\end{abstract}

\begin{keywords}
accretion, accretion discs -- X-rays: binaries -- X-Rays: galaxies -- X-rays: individuals (IC 342 X-1, NGC 5204 X-1, NGC 5408 X-1, Holmberg IX X-1, Holmberg II X-1 and NGC 253 X-1)
\end{keywords}

\section{Introduction}

Ultra Luminous X-ray sources (ULXs) are a peculiar class of extragalactic, point like and off-nuclear X-ray sources with isotropic luminosity in excess of $10^{39}$ erg s$^{-1}$ (e.g. \citealt{fengsoria11}). Although they were discovered more than 30 years ago and nowadays more than 450 ULXs are known and catalogued (e.g. \citealt{roberts00, colbert02, swartz04, liu05, walton11}), their nature is still matter of debate and their observational properties are still puzzling. A number of observational results, including the X-ray variability and the existence of periodic modulations in the X-ray or optical flux of some sources, suggest that they can be accreting black hole (BH) binary systems.
Their extreme luminosities may be explained in terms of sub-Eddington accretion onto Intermediate Mass BHs (IMBHs, {100-$10^4$ $M_{\odot}$}; \citealt{colbert99}), beamed and/or super-Eddington accretion onto stellar mass BHs ({5-20 $M_{\odot}$}; e.g. \citealt{king01,begelman02,fengsoria11}) or near-Eddington accretion onto massive stellar BHs formed from low metallicity stars ({30-80 $M_{\odot}$}, e.g. \citealt{mapelli09,zampieri09,belczynski10}). 

Based on the early, low counting statistics XMM-Newton spectra BH masses significantly in excess of 100 up to {\bf $\sim10^4$ $M_{\odot}$} were inferred \citep[e.g.][]{miller03,miller04}. However more recent high quality XMM-Newton observations have shown that the X-ray spectra of ULXs are characterised by properties not commonly observed in Galactic BH X-ray binary systems (XRBs) accreting at sub-Eddington rates (\citealt{stobbart06, goncalves06}). A roll-over at high energy, usually at 3-5 keV, is often observed coupled to a soft excess \citep{stobbart06}. \citet{gladstone09} found that it was almost ubiquitous in the highest quality ULX spectra and proposed
that such curvature is a characteristic feature of a new spectral state, the \textit{ultraluminous state} \citep{roberts07}. They modelled these spectra in terms of an optically thick corona coupled to an accretion disc. In particular, they proposed the existence of a spectral sequence in which the soft component becomes more and more important as the luminosity increases. {For the highest luminosity sources, this component was later interpreted as originating from outflow ejections from the disc (\citealt{middleton11}a)}.
This is consistent with recent theoretical calculations of the hydrodynamic structure of accretion discs at super-Eddington rates (e.g. \citealt{poutanen07,ohsuga07,ohsuga09}).

\begin{table*}
  \begin{center}
   \caption{Log of the observations of the ULXs analysed in this work.}
    \scalebox{0.9}{\begin{minipage}{15.0cm}
\footnotesize
   \label{ccd6}
   \begin{tabular}{l c c c c c c c  }
\hline
No. & Obs. ID & Date & Exp$^a$ & Instr. $^b$ & EPIC-pn  & Net counts & Off-axis angle\\
& & & (ks) & &count s$^{-1}$ && \\
\hline
\multicolumn{8}{c}{{\bf IC 342 X-1}; D=3.3 Mpc$^c$;  N$_{\rm H_{Gal}}$ = 31.1$\cdot10^{20}$ cm$^{-2}$   $^d$; Ra, Dec (J2000): 03 45 55.5, +68 04 54.2 }\\
\\
1 & 0093640901 & 2001-02-11 & 4.83 & pn & 0.37 & 1761 &5.08' \\
2 & 0206890101 & 2004-02-20 & 17.06 & pn/M1/M2 & 0.42 & 6863,3212,3176 &2.49' \\
3 & 0206890201 & 2004-08-17 & 14.36 & pn/M1/M2 & 0.90 & 12473,7757,7861 & 4.27'\\
4 & 0206890401 & 2005-02-10 & 5.92 & pn/M1/M2 & 1.11 & 6593,3718,4409 &	2.63' \\
\hline
\multicolumn{8}{c}{{\bf NGC 253 X-1}; D=3.9 Mpc$^e$;  N$_{\rm H_{Gal}}$ = 1.3$\cdot10^{20}$ cm$^{-2}$   $^d$; Ra, Dec (J2000): 00 47 22.56, -25 20 51.0 }\\
\\
1 & 0110900101 & 2000-12-13 & 10.97 & pn/M1/M2 & 0.105 & 1154,802, 893 & 4.57'\\
2 & 0152020101 & 2003-06-19 & 61.37 & pn/M1/M2 & 0.271 & 16625,6978,7201& 5.30' \\
3 & 0304851101 & 2005-02-10 & 10.93 & pn/M1/M2 & 0.097 & 1057,727,694 &3.13' \\
4 & 0304850901 & 2006-01-02 & 8.82 & pn/M1/M2 & 0.141 & 1241,458,488 &3.14'\\
5 & 0304851001 & 2006-01-06 & 8.75 & pn/M1/M2 & 0.155 & 1359,530,549 &3.17' \\
6 & 0304851201 & 2006-01-09 & 16.00 & pn/M1/M2 & 0.158 & 2525,994,965 &3.19' \\
7 & 0304851301 & 2006-01-11 & 4.34 & pn/M1/M2 & 0.162 & 703,338,373 &3.22' \\
\hline
\multicolumn{8}{c}{{\bf NGC 5204 X-1}; D=4.8 Mpc $^f$;  N$_{\rm H_{Gal}}$ = 1.39$\cdot10^{20}$ cm$^{-2}$   $^d$; Ra, Dec (J2000): 13 29 38.6, +58 25 05.7}\\
\\
1 & 0142770101 & 2003-01-06 & 15.33 & pn/M1/M2 & 0.624 & 9560,3121,3211 &1.14' \\
2 & 0142770301 & 2003-04-25 & 4.02 & pn/M1/M2 & 0.862 & 3465,1903,1868& 1.12'\\
3 & 0150650301 & 2003-05-01 & 5.26 & pn/M1/M2 & 1.018 & 5357,2213,2327 &1.30' \\
4 & 0405690101 & 2006-11-15 & 10.05 & pn/M1/M2 & 1.228 & 12600,7752,7752 &1.10' \\
5 & 0405690201 & 2006-11-19 & 31.25 & pn/M1/M2 & 1.031 & 32219,11791,11957 &1.08'\\
6 & 0405690501 & 2006-11-25 & 22.42 & pn/M1/M2 & 0.775 & 17364,6906,7174 &1.13' \\
\hline
\multicolumn{8}{c}{{\bf NGC 5408 X-1}; D=4.8 Mpc$^e$;  N$_{\rm H_{Gal}}$ = 5.67$\cdot10^{20}$ cm$^{-2}$   $^d$; Ra, Dec (J2000): 14 03 19.6, -41 22 59.6}\\
\\
1 & 0112290501 & 2001-07-31 & 3.68 & pn/M1/M2 & 1.441 & 5303,2530,2642 & 1.26' \\
2 & 0112290601 & 2001-08-08 & 4.50 & pn/M1/M2 & 1.337 & 6023,2024,2140 &1.27' \\
3 & 0112290701 & 2001-08-24 & 7.50 (MOS1) & M1/M2 & 1.018 (MOS1) & 2405,2482 &1.23' \\
4 & 0112291201 & 2003-01-27 & 2.79 & pn/M1/M2 & 1.228 & 2354,920,935 &0.99' \\
5 & 0302900101 & 2006-01-13 & 92.54 & pn/M1/M2 & 1.031 & 94298,25437,25331 &1.09' \\
6 & 0500750101 & 2008-01-13 & 46.09 & pn/M1/M2 & 0.775 & 43753,18471,17782 &1.07' \\
7 & 0653380201 & 2010-07-17 & 92.68 & pn/M1/M2 & 1.137 & 105377,26034,33154 &1.11' \\
8 & 0653380301 & 2010-07-19 & 96.86 & pn/M1/M2 & 1.113 & 107805,30145,30188 &1.12' \\
9 & 0653380401 & 2011-01-26 & 87.38 & pn/M1/M2 & 1.061 & 92710,29307,29126 &1.12' \\
10 & 0653380501 & 2011-01-28 & 88.51 & pn/M1/M2 & 1.024 & 90634,27952,27822&1.12' \\
\hline
\multicolumn{8}{c}{{\bf Ho II X-1}; D=4.5 Mpc$^e$;  N$_{\rm H_{Gal}}$ = 3.42$\cdot10^{20}$ cm$^{-2}$   $^d$; Ra, Dec (J2000): 08 19 29.0, +70 42 19.3}\\
\\
1 & 0112520601 & 2002-04-10 & 4.64 & pn/M1/M2 & 3.054 & 14164, 8316, 8886 &1.13' \\
2 & 0112520701 & 2002-04-16 & 3.77 & pn/M1/M2 & 2.751 & 10377, 4796, 4994&1.11' \\
3 & 0112520901 & 2002-09-18 & 4.33 & pn/M1/M2 & 0.815 & 3527, 1317, 1434&1.11' \\
4 & 0200470101 & 2004-04-15 & 40.75 & pn/M1/M2 & 3.056 & 124532, 49487, 50578&1.14'\\
5 & 0561580401 & 2010-03-26 & 23.19 & pn/M1/M2 & 1.238 & 28709, 14098, 13993&1.14' \\
\hline
\multicolumn{8}{c}{{\bf Ho IX X-1}; D=3.55 Mpc$^g$;  N$_{\rm H_{Gal}}$ = 4.06$\cdot10^{20}$ cm$^{-2}$   $^d$; Ra, Dec (J2000): 09 57 53.2, +69 03 48.3}\\
\\
1 & 0112521001 & 2002-04-10 & 7.05 & pn/M1/M2 & 1.895 & 13350,5806,5782 &1.11'\\
2 & 0112521101 & 2002-04-16 & 7.64 & pn/M1/M2 & 2.173 & 16610,7041,7338&1.13' \\
3 & 0200980101 & 2004-09-26 & 83.17 & pn/M1/M2 & 1.495 & 124339,46782,47339&1.13' \\
4 & 0657801601 & 2011-04-17 & 0.96 & pn/M1/M2 & 0.760 & 733,2851,2807&7.40' \\
5 & 0657801801 & 2011-09-26 & 7.40 & pn/M1/M2 & 2.477 & 13830,11969,13487 &5.29' \\
6 & 0657802001 & 	2011-03-24 & 3.20 & pn/M1/M2 & 1.369 & 4381,2882,3672&7.32' \\
7 & 0657802201 & 2011-11-23 & 13.10 & pn/M1/M2 & 2.211 & 28964,16186,15303&5.21'\\
\hline
\end{tabular}
\end{minipage}}
\end{center}
$^a$ GTI of EPIC-pn; $^b$ pn = EPIC-pn camera; M1/M2 = EPIC-MOS1/MOS2 camera; $^c$ \citet{saha02}; $^d$ \citet{dickey90}; $^e$ \citet{karachentsev03}; $^f$ \citet{stobbart06}; $^g$ \citet{freedman94}; 
\end{table*}

The increasing number of observations of ULXs that became available in the last years also prompted an investigation of their long-term spectral state variability. 
IC 342 X-1 and X-2 were the first two ULXs in which transitions from a low/hard to a high/soft spectral state were observed \citep{kubota01}. However, the behaviour of the soft component is intriguing. In fact, although IC 342 X-2 shows a correlation between the disc luminosity and temperature typical of a standard accretion disc, IC 342 X-1 shows the opposite behaviour \citep{feng09}. An anti-correlation was observed also in other sources \citep{kajava09}. Changes reminiscent of Galactic XRB spectral transitions were also observed in NGC 1313 X-1 and X-2 \citep{feng06}. A systematic analysis of their X-ray spectral variability in terms of a comptonisation model plus a multicolor blackbody disc was performed by \citet{pintore12}, who characterised the behaviour of these two ULXs in terms of variations in the optical thickness of the Comptonizing corona (\textit{very thick} and \textit{thick} state). The resulting picture is that ULXs show rather peculiar spectral changes, often different from source to source.

As shown by at least 30 years of studies of Galactic XRBs, in addition to the spectral shape, short term variability is very important to understand their accretion states (e.g. \citealt{belloni10}). Few sources (NGC 5408 X-1, M82 X-1 and X42.3+59) show quasi periodic oscillations (QPOs), the classification of which remains still unclear (e.g. \citealt{strohmayer03b,mucciarelli06,strohmayer07,strohmayer09,feng10,middleton11b}b; \citealt{dheeraj12, caballero13}).
In general, the properties of the short term variability of ULXs are still poorly understood. In fact, sources with similar X-ray spectra show different temporal variability. \citet{heil09} analysed the Power Spectral Densities (PSD) of a sample of 16 bright ULXs and found that, irrespectively of their X-ray spectra, there are two groups of sources: a smaller group displays a well defined variability at about the same level ($\sim10\%$), while in the other one the variability is almost absent. Recently, \citet[][a,b]{middleton11} showed that also the analysis of the energy dependence of the short term variability can be a powerful tool to discriminate among different spectral models. In particular, they suggested the possibility that the short-term variability is produced by turbulences in a clumpy wind, which from time to time encounters our line of sight.

The aim of the present work is to investigate in a systematic way the spectral variability of ULXs on a sample of sources selected to have high quality {\textit{XMM-Newton} observations. We complement the spectral analysis with a careful investigation of their short term variability. We attempt to characterize the spectral variability using also the \textit{hardness ratios} and \textit{colour-colour diagrams}, that have been successfully adopted in the past to study the behaviour of XRBs. 
Considering the relevance of the metallicity for some of the proposed scenarios for the formation of ULXs, we also put a certain effort in using the highest counting statistics X-ray spectra to attempt a measurement of the chemical abundances in the environment of ULXs, following the approach of \citet{winter07} and \citet{pintore12}.
{A systematic analysis of the spectral and temporal variability of a larger sample of ULXs has very recently been published also by \citet{sutton13}. {Adopting a multicolour disc plus a powerlaw, they classified the ULXs in three spectral regimes, defined as \textit{broadened disc, hard ultraluminous} and \textit{soft ultraluminous}. The first is characterised by a disc-like spectral shape while the other two by the predominance of either the powerlaw component at high energies or the disc component at low energies.} While in some respects their investigation is complementary to ours (the goal and some of the results are similar to those reported here), {in others our approach differs because we did not use fluxes from a spectral model for computing hardness-intensity and colour-colour diagrams, but adopted the \textit{XMM-Newton} counts in different energy bands. We will show that this method allows us to classify ULXs also in case of low quality data and weak constraints on the parameters of the spectral models, and therefore can be applied to a larger sample of sources.}}

The plan of the paper is the following. In Section~\ref{selection} we describe the selection of the sample of ULXs. In Section~\ref{data_reduction} we present the adopted X-ray data reduction procedure and the result of the X-ray spectral and temporal analysis of all the sources. In Section~\ref{colorrr}, we tentatively try to interpret the spectral evolution of ULXs on the \textit{colour-colour} and \textit{hardness-intensity} diagrams and, finally, we discuss our results in Section~\ref{discussion}.

\section{Selection of the sample}
\label{selection}

We selected a sample of ULXs from the catalogues of \citet{liu05} and \citet{walton11} that were observed by \textit{\textit{XMM-Newton}}. In order to detect the high energy curvature in the EPIC spectrum, at least 10000 counts are needed \citep{gladstone09}. In addition, a high number of total counts is also requested to perform an analysis of the chemical abundances. \citet{winter06} showed that the Oxygen or Iron K-shell edges can be detected with at least 5000 or 40000 counts in the EPIC instrument, respectively. The observed count rates of the closest ULXs in the EPIC-\textit{pn} detector are in the range $\sim 0.1-1.5$ count s$^{-1}$ and hence long exposure times of $\sim 15-100$ ks are needed to provide the required amount of total counts.

Following these constraints we select nearby ULXs (D $\leq$ 5 Mpc) that have at least one long observation with \textit{\textit{XMM-Newton}} (t$_{exp} \geq $15 ks). {This gives us the required $\sim$10000 counts in the EPIC instrument}. The additional requirement of at least three  \textit{\textit{XMM-Newton}} datasets in different epochs (even if {some epochs are of shorter exposures}) allows us to study the spectral variability. The final list includes:  IC 342 X-1, NGC 253 X-1, NGC 5204 X-1, NGC 5408 X-1, Ho IX X-1 and Ho II X-1, NGC 1313 X-1 and X-2.
The sources in NGC 1313 have been presented in \citet{pintore12} and will be added to the discussion of the results. This sample might not be fully representative of the ULX properties but the adopted selection criteria allow us to sample a rather large range of luminosities (L$_X \sim 1-30 \times 10^{39}$ erg s$^{-1}$).

\section{Data Analysis of spectral and timing properties}
\label{data_reduction}

\subsection{Data Reduction}
\label{data_reduction1}

We carried out a complete spectral and temporal analysis on all the available \textit{XMM-Newton} observations of the ULXs listed above. We excluded from the analysis observations performed earlier than December 2000, because the calibration before this date may be incomplete. Data were reduced using SAS v. 11.0.0, extracting spectra and lightcurves from events with {\sc pattern}$\leq 4$ for EPIC-pn (which allows for single and double pixel events) and {\sc pattern}$ \leq 12$ for EPIC-MOS (which allows for single, double, triple and quadruple pixel events). We set `{\sc flag}=0' in order to exclude bad pixels and events coming from the CCD edge. We also excluded from the analysis the observations in which the sources were in the CCD gap and a number of observations severely affected by strong solar flares. In order to avoid distortions induced by background particles, EPIC-MOS and EPIC-pn spectra were extracted selecting good time intervals with a background count rate in the entire field of view not higher than 0.7 count s$^{-1}$ in the 10 $-$ 12 keV energy range. A few observations of NGC 5408 X-1 and Ho II X-1 were excluded from the analysis because the event list was empty. The log of observations and relevant information on the sources, including total net counts in the various instruments, is reported in Table~\ref{ccd6}.

Spectra and lightcurves of IC 342 X-1, NGC 253 X-1, Ho II X-1, Ho IX X-1 and NGC 5408 X-1 were obtained selecting circular extraction regions of 30$"$ and 65$"$ for source and background (when possible, on the same CCD where the source is located), respectively.
NGC 5204 X-1 was often very close to the CCD gap and hence the extraction region was different from observation to observation (ranging from $24"$ to 31$"$ {for the source and from $50"$ to 65$"$ for the background}). 

All the spectra were rebinned with 25 counts per bin in order to apply the $\chi^2$ statistics. The spectral fits were performed using XSPEC v. 12.6.0 \citep{arnaud96}. To improve the counting statistics we fitted EPIC-pn and EPIC-MOS spectra simultaneously in the 0.3-10 keV energy range.
In all fits, a multiplicative constant was introduced for the three instruments to account for possible residual differences in calibration. The EPIC-pn constant was fixed to 1, while for the MOS they were allowed to vary. In general, the difference among the three instruments was less than 10$\%$.

\subsection{Spectral fits}
\label{spectral_fits}

\begin{table*}
\footnotesize
\begin{center}
   \caption{Best fitting spectral parameters obtained with the absorbed \textit{diskbb+comptt} model. The errors are at 90$\%$ for each parameter of interest.}
   \label{ic342_table}
   \begin{tabular}{l c c c c c c c c}
\hline
No. &  $N_H$$^a$ & $kT_{disc}$$^{b,c}$ & $kT_{cor}$$^d$ & $\tau$$^e$ & APEC$^f$ & $L_{X}$ [0.3-10 keV] $^g$ & $L_{disc}$ [0.3-10 keV]$^h$ & $\chi^2/dof$ \\ 
& ($10^{21}$ cm$^2$) & (keV) & (keV) & & keV & ($10^{39}$ erg s$^{-1}$) & ($10^{39}$ erg s$^{-1}$) & \\
\hline
\multicolumn{9}{c}{IC 342 X-1}  \\ 
\hline
1&  $6.7_{-0.5}^{+0.5}$	& $0.213_{-0.007}^{+0.007}$ & $3.2_{-1.2}^{+0.1}$ & $6.6_{-2}^{+2}$ & & $5.6_{-1.1}^{+1.2}$ & $1.2_{-0.4}^{+0.6}$ & 57.17/58\\
2& $5.5_{-0.2}^{+0.2}$	& $0.390_{-0.003}^{+0.003}$ & $2.21_{-0.04}^{+0.04}$ & $9.0_{-0.2}^{+0.2}$ & & $5.7_{-0.5}^{+0.5}$ & $1.7_{-0.3}^{+0.2}$  & 443.53/429\\
3& $6.2_{-0.1}^{+0.1}$ & $ 0.494_{-0.003}^{+0.003}$ & $1.73_{-0.02}^{+0.02}$ & $9.9_{-0.1}^{+0.1}$ & & $11.3_{-1.1}^{+0.4}$ & $3.8_{-0.3}^{+0.2}$ & 753.10/788\\ 
4& $5.3_{-0.2}^{+0.2}$     &  $0.348_{-0.007}^{+0.007}$ & $1.98_{-0.02}^{+0.02}$ & $8.5_{-0.1}^{+0.1}$ & & $13.3_{-1.1}^{+1.2}$ & $1.1_{-0.2}^{+0.3}$ & 479.56/485\\  
\hline
\hline
\multicolumn{9}{c}{NGC 253 X-1} \\ 
\hline
1& $0.97_{-0.01}^{+0.01}$	& $0.236_{-0.004}^{+0.004}$ & $2.8_{-0.1}^{+0.1}$ & $5.6_{-0.3}^{+0.4}$ & & $0.8_{-0.2}^{+0.2}$ & $0.3_{-0.05}^{+0.04}$  & 93.88/97\\
2& $0.82_{-0.04}^{+0.04}$     &  $0.794_{-0.002}^{+0.002}$ & $1.25_{-0.03}^{+0.03}$ & $27_{-4}^{+5}$ & & $2.4_{-0.1}^{+0.1}$ & $1.7_{-0.5}^{-0.2}$ & 852.44/810\\  
3& $0.64_{-0.01}^{+0.01}$	&  $0.104_{-0.007}^{+0.007}$ & $0.94_{-0.02}^{+0.02}$ & $13.8_{-0.4}^{+0.4}$ & & $0.72_{-0.14}^{+0.19}$ & $<0.26$  & 88.57/91\\
4& $0.20_{-0.02}^{+0.02}$	&  $0.464_{-0.009}^{+0.009}$ & $0.97_{-0.04}^{+0.04}$ & $20_{-2}^{+3}$ & & $0.99_{-0.11}^{+0.31}$ & $0.33_{-0.07}^{+0.11}$ &  75.117/73\\ 
5& $0.45_{-0.02}^{+0.02}$ & $ 0.35_{-0.01}^{+0.01}$ & $1.62_{-0.06}^{+0.06}$ & $9.5_{-0.6}^{+0.6}$ & & $1.2_{-0.23}^{+0.3}$ & $0.21_{-0.06}^{+0.06}$ & 88.229/83\\
6& $0.49_{-0.01}^{+0.01}$	&  $0.715_{-0.008}^{+0.008}$ & $1.61_{-0.08}^{+0.08}$ & $10_{-1}^{+1}$ & & $1.30_{-0.2}^{+0.3}$ & $0.64_{-0.07}^{+0.08}$ & 174.57/161\\
7& $0.40_{-0.03}^{+0.03}$ & $0.79_{-0.002}^{+0.002}$ & $1.80_{-0.02}^{+0.02}$ & $9_{-3}^{+3}$ & & $1.30_{-0.44}^{+0.5}$ & $0.68_{-0.13}^{+0.16}$ & 44.325/45\\ 
\hline
\hline
\multicolumn{9}{c}{Ho IX X-1} \\ 
\hline
1& $1.26_{-0.05}^{+0.05}$ & $0.240_{-0.002}^{+0.002}$ & $2.75_{-0.03}^{+0.03}$ & $7.2_{-0.1}^{+0.1}$ & & $13.8_{-0.9}^{+1}$ & $2.0_{-0.2}^{+0.2}$ & 685.17/706\\
2& $1.23_{-0.05}^{+0.05}$ & $ 0.219_{-0.002}^{+0.002}$ & $2.85_{-0.03}^{+0.03}$ & $6.66_{-0.08}^{+0.08}$ & & $15.8_{-0.9}^{+1}$ & $1.3_{-0.1}^{+0.2}$ & 749.82/817\\ 
3& $1.35_{-0.02}^{+0.02}$	& $0.245_{-0.001}^{+0.001}$ & $2.38_{-0.01}^{+0.01}$ & $9.07_{-0.05}^{+0.05}$ & & $12.1_{-0.3}^{+0.3}$ & $2.19_{-0.05}^{+0.06}$  & 2182.62/2044\\
4& $0.60_{-0.1}^{+0.1}$     &  $0.320_{-0.010}^{+0.010}$ & $3.03_{-0.07}^{+0.08}$ & $6.4_{-0.2}^{+0.2}$ & & $22_{-2}^{+2}$ & $2.3_{-0.5}^{+0.5}$ & 159.82/195\\  
5& $1.09_{-0.04}^{+0.04}$	&  $0.258_{-0.003}^{+0.003}$ & $1.91_{-0.01}^{+0.01}$ & $8.45_{-0.08}^{+0.08}$ & & $23_{-2}^{+1}$ & $1.7_{-0.1}^{+0.2}$ &  907.47/967\\ 
6& $1.60_{-0.08}^{+0.08}$ & $ 0.250_{-0.002}^{+0.002}$ & $4.5_{-0.1}^{+0.1}$ & $6.4_{-0.2}^{+0.2}$ & & $16_{-1}^{+2}$ & $2.9_{-0.3}^{+0.4}$ & 340.03/350\\
7& $1.22_{-0.04}^{+0.04}$ & $ 0.266_{-0.002}^{+0.002}$ & $1.86_{-0.01}^{+0.01}$ & $8.74_{-0.07}^{+0.07}$ & & $28_{-1}^{+2}$ & $2.2_{-0.2}^{+0.2}$ & 1122.67/1154\\
\hline
\hline
\multicolumn{9}{c}{NGC 5204 X-1} \\ 
\hline
1&$0.04_{-0.04}^{+0.04}$ & $0.306_{-0.002}^{+0.002}$ & $1.39_{-0.02}^{+0.02}$ & $12.7_{-0.4}^{+0.4}$ &  & $5.1_{-0.3}^{+0.3}$ & $1.90_{-0.4}^{+1.2}$ & 398.45/453\\
2& $0.67_{-0.06}^{+0.06}$ & $ 0.203_{-0.002}^{+0.002}$ & $6.20_{-0.2}^{+0.2}$ & $3.5_{-0.1}^{+0.2}$ & & $7.7_{-0.9}^{+1}$ & $2.3_{-0.3}^{+0.3}$ & 238.65/221\\ 
3& $0.31_{-0.05}^{+0.05}$	& $0.283_{-0.002}^{+0.002}$ & $1.55_{-0.03}^{+0.03}$ & $8.8_{-0.3}^{+0.3}$ & & $7.7_{-0.8}^{+0.9}$ & $3.1_{-0.2}^{+0.3}$  & 301.49/271\\
4& $0.33_{-0.03}^{+0.03}$     &  $0.216_{-0.001}^{+0.001}$ & $1.44_{-0.01}^{+0.01}$ & $7.2_{-0.1}^{+0.1}$ & & $8.7_{-0.6}^{+0.6}$ & $2.3_{-0.2}^{+0.2}$ & 717.93/722\\  
5& $0.47_{-0.02}^{+0.02}$	&  $0.265_{-0.001}^{+0.001}$ & $2.35_{-0.02}^{+0.02}$ & $5.72_{-0.02}^{+0.08}$ & & $8.1_{-0.4}^{+0.4}$ & $2.9_{-0.1}^{+0.1}$ &  838.98/822\\ 
6& $0.32_{-0.03}^{+0.03}$ & $ 0.231_{-0.001}^{+0.001}$ & $4.31_{-0.05}^{+0.05}$ & $4.6_{-0.05}^{+0.09}$ & & $6.6_{-0.4}^{+0.4}$ & $1.8_{-0.1}^{+0.1}$ & 642.82/676\\
\hline
\hline
\multicolumn{9}{c}{NGC 5408 X-1}  \\ 
\hline
1& $0.11_{-0.03}^{+0.03}$	& $0.180_{-0.001}^{+0.001}$ & $2.00_{-0.04}^{+0.04}$ & $5.0_{-0.2}^{+0.2}$ & 0 & $10.0_{-0.5}^{+0.2}$ & $6.6_{-0.4}^{+0.5}$ & 280.79/271\\
2 & $0.09_{-0.03}^{+0.03}$ & $ 0.192_{-0.001}^{+0.001}$ & $84_{-2}^{+2}$ & $0.11_{-0.01}^{+0.02}$ &0 & $9.8_{-0.9}^{+1}$ & $6.2_{-0.5}^{+0.5}$ & 235.86/273\\ 
3 & $0.28_{-0.05}^{+0.05}$	& $0.177_{-0.001}^{+0.001}$ & $1.03_{-0.03}^{+0.03}$ & $8.9_{-0.3}^{+0.3}$ &0 & $11.0_{-0.1}^{+0.1}$ & $7.4_{-0.6}^{+0.7}$  & 93.49/133\\
4& $0.72_{-0.07}^{+0.07}$     &  $0.171_{-0.001}^{+0.001}$ & $1.56_{-0.04}^{+0.04}$ & $6.8_{-0.3}^{+0.3}$ &0 & $8.2_{-1.2}^{+1.3}$ & $4.7_{+0.6}^{-0.6}$ & 118.3/127\\  
5& $0.62_{-0.01}^{+0.01}$	&  $0.153_{-0.002}^{+0.002}$ & $1.656_{-0.008}^{+0.008}$ & $6.04_{-0.04}^{+0.04}$ &$0.92_{-0.06}^{+0.06}$ & $8.6_{-0.2}^{+0.2}$ & $4.5_{-0.2}^{+0.1}$ &  968.60/913\\ 
6 & $0.61_{-0.01}^{+0.01}$ & $ 0.154_{-0.003}^{+0.003}$ & $1.506_{-0.009}^{+0.009}$ & $6.77_{-0.05}^{+0.05}$ & $0.84_{-0.03}^{+0.04}$& $7.9_{-0.3}^{+0.3}$ & $3.7_{-0.1}^{+0.1}$ & 868.60/796\\
7& $0.59_{-0.01}^{+0.01}$ & $ 0.156_{-0.002}^{+0.002}$ & $1.813_{-0.008}^{+ 0.008}$ & $5.88_{-0.03}^{+ 0.03}$ & $0.95_{-0.03}^{+0.03}$& $9.6_{-0.2}^{+0.2}$ & $4.3_{-0.1}^{+0.1}$ & 1085.20/1030\\
8 & $0.43_{-0.01}^{+0.01}$ & $ 0.163_{-0.001}^{+ 0.001}$ & $1.882_{-0.008}^{+ 0.008}$ & $5.73_{-0.03}^{+ 0.03}$ & $0.96_{-0.04}^{+0.04}$& $8.9_{-0.2}^{+ 0.3}$ & $3.8_{-0.1}^{+0.1}$ & 1158.99/1023\\
9 & $0.64_{-0.01}^{+0.01}$ & $ 0.154_{-0.001}^{+ 0.001}$ & $1.728_{-0.007}^{+ 0.007}$ & $5.97_{-0.03}^{+ 0.03}$ & $0.902_{-0.04}^{+0.04}$& $9.1_{-0.3}^{+ 0.2}$ & $4.2_{-0.1}^{+0.1}$ & 1045.72/981\\
10 & $0.55_{-0.01}^{+0.01}$ & $ 0.160_{-0.03}^{+ 0.04}$ & $2.047_{-0.009}^{+ 0.009}$ & $5.43_{-0.03}^{+ 0.03}$ & $0.87_{-0.03}^{+0.04}$& $8.5_{-0.3}^{+ 0.6}$ & $3.8_{-0.1}^{+0.1}$ & 1055.82/995\\
\hline
\hline
\multicolumn{9}{c}{Ho II X-1}  \\ 
\hline
1& $0.39_{-0.03}^{+0.03}$ & $0.189_{-0.001}^{+0.001}$ & $4.69_{-0.04}^{+0.04}$ & $3.10_{-0.04}^{+0.04}$     & & $21_{-1}^{+1}$ & $3.2_{-0.3}^{+0.3}$ & 576.28/629\\
2& $0.77_{-0.04}^{+0.04}$ & $ 0.200_{-0.004}^{+0.004}$& $2.16_{-0.03}^{+0.03}$ & $6.2_{-0.1}^{+0.1}$ 	      & & $22_{-1}^{+1}$     & $7.4_{-0.3}^{+0.3}$    & 507.21/532\\ 
3& $0.75_{-0.06}^{+0.06}$  & $0.167_{-0.001}^{+0.001}$ & $1.00_{-0.02}^{+0.02}$ & $8.4_{-0.2}^{+0.2}$        & & $5.7_{-0.7}^{+0.5}$ & $2.5_{-0.3}^{+0.3}$  & 191.13/185\\
4& $0.48_{-0.01}^{+0.01}$ &  $0.203_{-0.001}^{+0.001}$& $2.63_{-0.01}^{+0.01}$ & $4.66_{-0.02}^{+0.02}$  & & $23_{-1}^{+1}$ & $5.5_{-0.2}^{+0.1}$   & 1326.82/1318\\  
5& $0.57_{-0.02}^{+0.02}$ &  $0.200_{-0.001}^{+0.001}$& $1.51_{-0.01}^{+0.01}$ & $7.17_{-0.07}^{+0.07}$ & & $8.6_{-0.3}^{+0.2}$ & $3.6_{-0.1}^{+0.2}$    &  819.61/760\\ 
\hline

   \end{tabular} 
 \end{center}
$^a$ Intrinsic column density in excess the Galactic one; $^b$ Inner disc temperature; $^c$ The seed photons temperature $T_0$ is assumed to be equal to $T_{disc}$; $^d$ Temperature of the corona; $^e$ Optical depth of the corona; $^f$ Temperature of the plasma component; $^g$ Unabsorbed total X-ray luminosity;  $^h$ Unabsorbed disc luminosity. \\
\end{table*}

\begin{figure*}
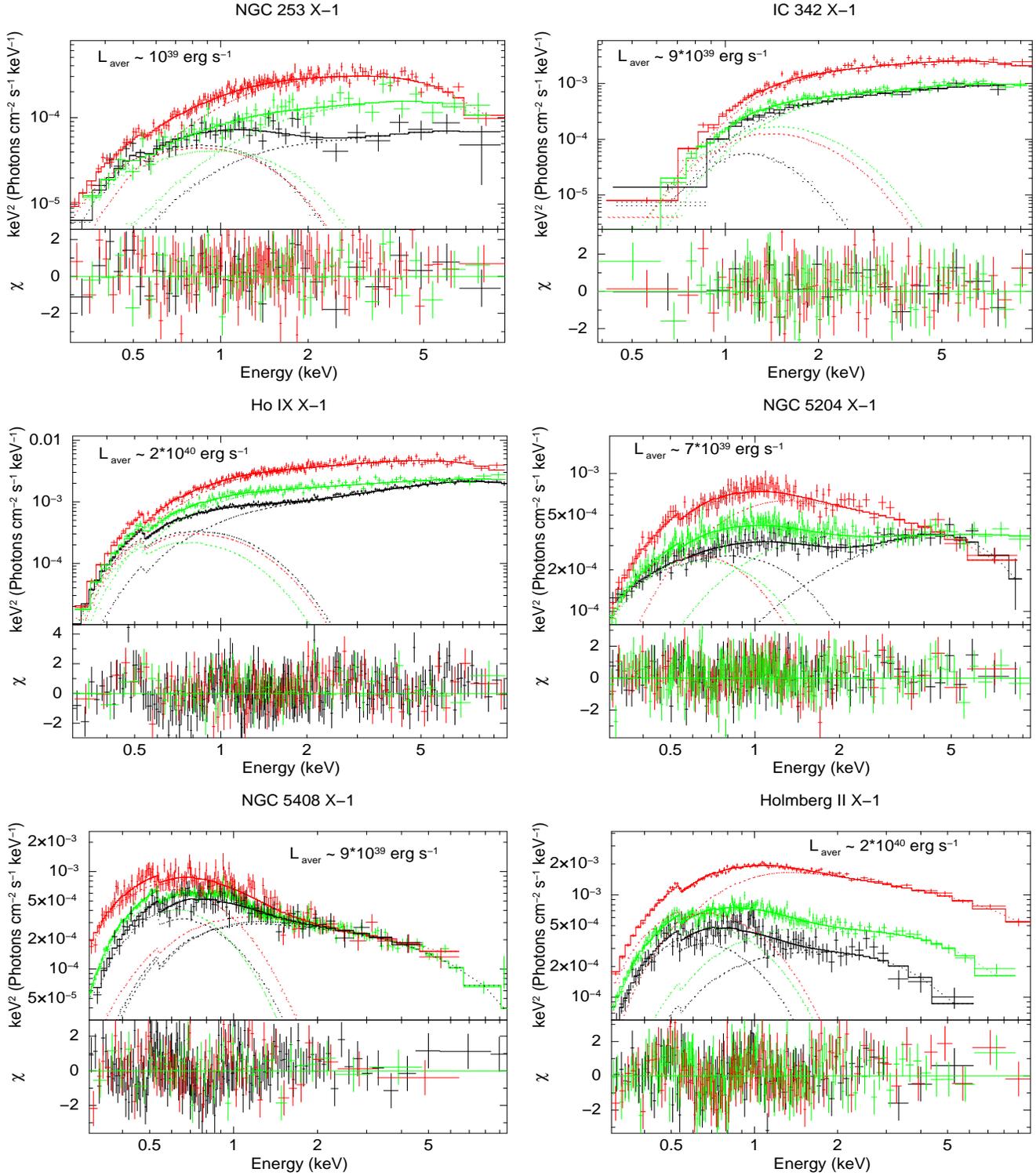

\center

\subfigure{\includegraphics[width=6.5cm,height=8.5cm,angle=-90]{2-2-2.ps}}
\hspace{3mm} \subfigure{\includegraphics[width=6.5cm,height=8.5cm,angle=-90]{1-2.ps}}
\subfigure{\includegraphics[width=6.5cm,height=8.5cm,angle=-90]{6-2.ps}}
\hspace{3mm}\subfigure{\includegraphics[width=6.5cm,height=8.5cm,angle=-90]{4-2.ps}}
\subfigure{\includegraphics[width=6.5cm,height=8.5cm,angle=-90]{5-2-2.ps}}
\hspace{3mm}\subfigure{\includegraphics[width=6.5cm,height=8.5cm,angle=-90]{3-2-2.ps}}

\caption{{Comparison of EPIC-pn unfolded ($E^2f(E)$) spectra, fitted with the \textit{diskbb+comptt} model (Table~\ref{ic342_table}). For display purposes, the spectra and the residuals were rebinned at a minimum of 5 $\sigma$. Only the highest (\textit{red}), lowest (\textit{black}) and medium luminosity (\textit{green}) spectra for each source are shown for clarity. The dashed lines are the \textit{diskbb} and \textit{comptt} components, while the solid line is the sum of the two (in NGC 5408, the APEC component is taken into account {for the fit} but not shown in the plot). {Going from \textit{top-left} to \textit{bottom-right}, the sources are ordered according to their position along the sequence on the hardness-intensity diagram of Figure~\ref{hardness_rebin} (see Section~\ref{HID}). The legend shows the average (0.3-10 keV) unabsorbed luminosity.} 
There is a trend in which the relative importance of the soft component increases with the luminosity of the source.
}}

\label{fig_confronto_IC342}
\end{figure*}

A combination of a multicolor disc plus a comptonisation model has been shown to offer a reasonable description of both poor and high quality ULX spectra and to provide a better fit than simple models as a multicolour blackbody disc (\textit{diskbb} in XSPEC, \citealt{mitsuda84}), \textit{powerlaw}, a slim disc (\textit{diskpbb} in XSPEC, \citealt{mineshige94}) or a \textit{diskbb+powerlaw} (i.e. \citealt{stobbart06,gladstone09,middleton11}a; \citealt{pintore12}).
Following these findings and guided by the idea of describing the spectral evolution of ULXs within a common framework, we then fitted the spectra of our sample of ULXs adopting as reference model a multicolor accretion disc (\textit{diskbb} in XSPEC) plus a comptonising component (\textit{comptt} in XSPEC; \citealt{titarchuk94}). 
We note that for some sources the spectral fits with this two-component model
often display several local minima with very similar values
of the $\chi^2$, sometimes with evidence for both a strong/warm and a weak/cool (or no) disc.
In Table~\ref{ic342_table} we quote the values
of the parameters and formal statistical errors {(at 90\% confidence level)} for the
absolute minima found with this model, 
but we stress that, for some poor quality spectra, the actual uncertainty on the spectral parameters caused by the topology of the $\chi^2$ surface may be larger.
{Because of the low statistics, in the majority of the cases we are not able to independently vary the temperature of the disc ($T_{disc}$) and that of the seed photons ($T_{0}$) and hence we tie them together.}
Although this is not fully consistent from a physical point of view (the corona is often optically thick and then the inner disc is hidden), leaving them free to vary independently within a factor of a few does not lead to significant qualitative changes in the results (see also \citealt{pintore12}).

Two absorption components (\textit{tbabs}, {\citealt{wilms00}, in XSPEC}) were considered for all spectral models: one fixed at the Galactic column density along the direction of the source and the second one, free to vary in order to account for local absorption. The adopted Galactic $N_H$ and distances are listed in Table~\ref{ccd6}, in the headers for each source. 
{In Figure~\ref{fig_confronto_IC342} we show the results of the spectral fits obtained at the lowest, medium and highest luminosity level for each source. They give a visual impression of the observed spectral and intensity variability. {The sources are ordered according to their position along the sequence on the hardness-intensity diagram of Figure~\ref{hardness_rebin}-\textit{left} (see Section~\ref{HID}). In general, the sources show a trend in which the relative importance of the soft component increases with the 
luminosity of the source.} In Table~\ref{ic342_table} we report the best fit parameters. For the highest quality observations of NGC 5408 X-1 we add an underlying plasma component (\textit{APEC} in XSPEC) with a temperature of $\sim0.9$ keV that leads to a significant improvement in the fit. This emission could be produced either by the diffuse gas in the host galaxy or directly from the environment around the source (\citealt{strohmayer07,middleton11b}b).

The (unabsorbed) luminosities of the sources of the sample span from $\sim 7\cdot 10^{38}$ erg s$^{-1}$ to $\sim 3\cdot10^{40}$ erg s$^{-1}$ (Figure~\ref{lightcurve_tutti}-\textit{top}), and therefore sample a wide range of observed ULX luminosities. 
Most of the sources exhibit intermediate luminosities, clustering around $\sim (7-8)\cdot10^{39}$ erg s$^{-1}$, but with significant long-term variability, which in some cases is higher than a factor of 3 (i.e. Holmberg IX X-1, NGC 1313 X-2 and IC 342 X-1).

In all the observations the comptonizing medium (corona) is optically thick and cold (Figure~\ref{lightcurve_tutti}-\textit{bottom}). 
The temperatures and optical depths are in the range $kT_{cor}\sim 1-6$ keV and $\tau \sim 3-30$, significantly lower the former and higher the latter than those seen in Galactic XRBs ($kT_{cor}\geq50$ keV and $\tau\leq1$, e.g. \citealt{mcclintock06})\footnote{We note that, for temperatures of the corona $\ga 2$ keV, the trend in the $kT_{cor}-\tau$ plane (Figure~\ref{lightcurve_tutti}-\textit{bottom}) is not very different from that expected for constant spectral index/Compton parameter and may then reflect in part some degeneracy in the data. However, below $\simeq 2$ keV, the behaviour is real because $kT_{cor}$ is well constrained within the {\it XMM-Newton} bandpass.}. This is consistent with the findings of \citet{gladstone09}, who interpret the \textit{ultraluminous state} in terms of an optically thick, cold corona physically coupled to the inner regions of an accretion disc. However, the fit of observation \#2 of NGC 5408 X-1 shows a more pronounced degeneracy in the parameters of the corona which is also consistent with being warm (84 keV) and marginally optically thin ($\tau\sim0.9$).
For this reason, this observation is not shown in the plot $\tau-kT_{cor}$ (Figure~\ref{lightcurve_tutti}-\textit{bottom}). 
 \begin{figure}
\center
\includegraphics[height=7.0cm,width=9.0cm]{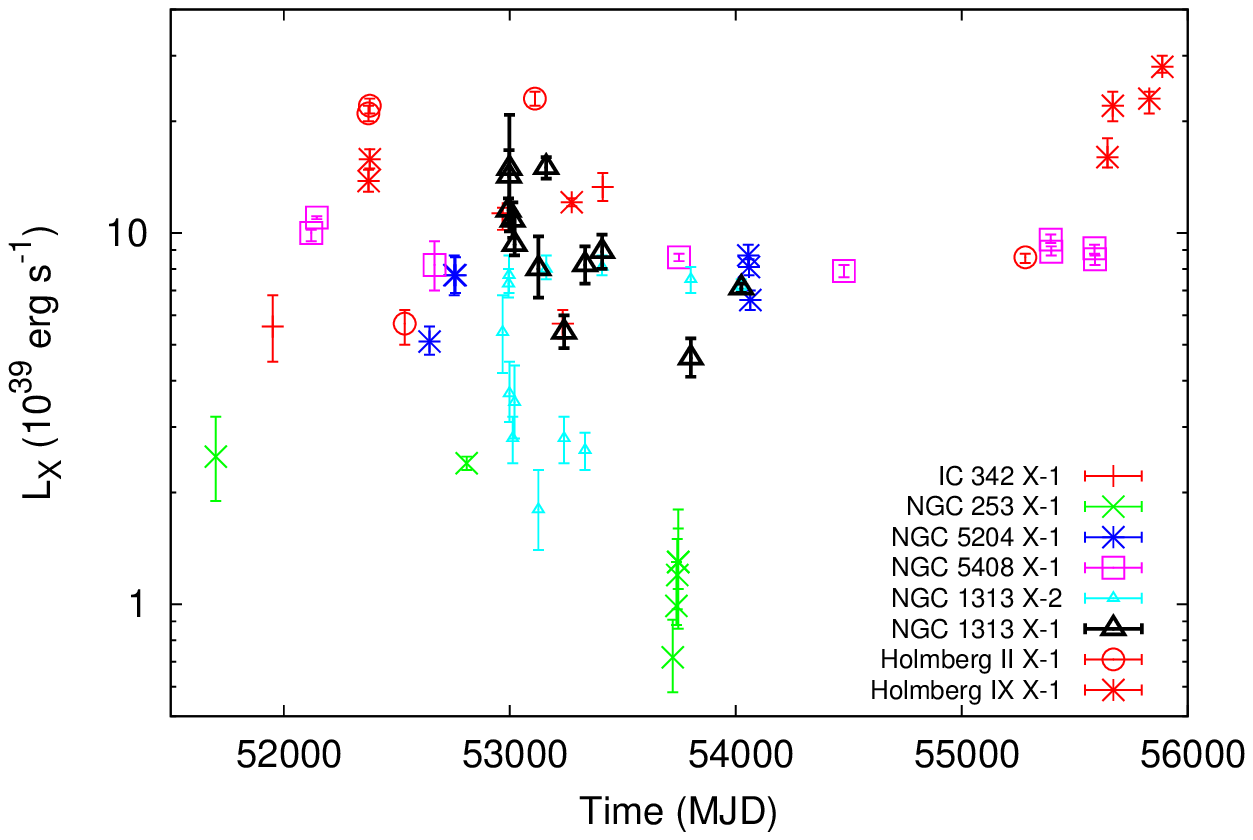}
\includegraphics[height=7cm,width=9cm]{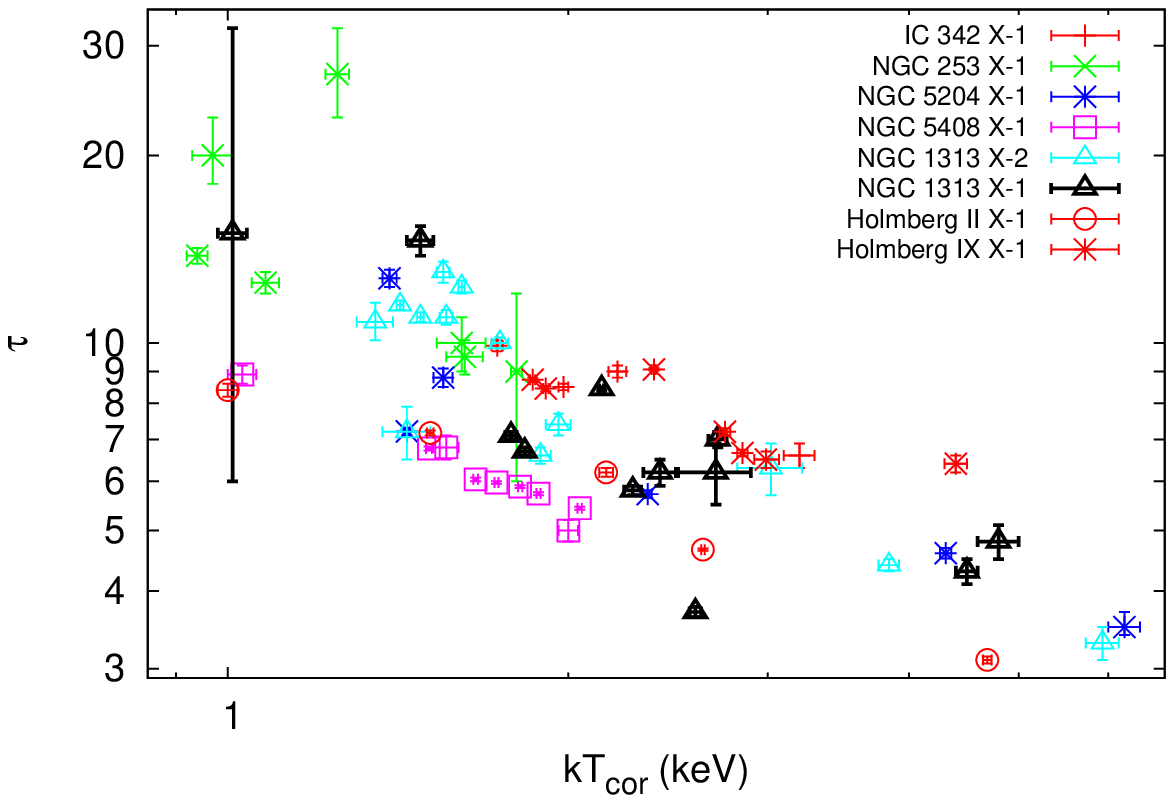}
\caption{\textit{top}: Unabsorbed total luminosities in the 0.3-10 keV energy band {as a function of time}; \textit{bottom}: optical depth $\tau$ versus temperature of the corona $kT_{cor}$ (\textit{diskbb+comptt} model). We added also NGC 1313 X-1 and X-2 for comparison (see \citealt{pintore12}). Different colors and symbols refer to the sources, as listed inset.}
\label{lightcurve_tutti}
\end{figure}

At such high optical depths and low temperatures, the physical conditions in the corona are rather different from those in Galactic XRBs.
{If the gas is pure hydrogen, the bremsstrahlung luminosity of the corona is $L_{brem,cor}=2.5\cdot10^{21}\cdot T_{cor}^{1/2}\cdot \tau_{es}\cdot r_{cor}$ erg s$^{-1}$, where $\tau_{es}$ is the electron-scattering optical depth and $r_{cor}$ is the radius of the corona. Assuming that $L_{brem,cor}$ is negligible ($<10^{38}$ erg s$^{-1}$)}, we find $r_{cor}\la 7\times 10^{10}$ cm for $kT_{cor}\approx 1$ keV and $\tau_{es} \approx 5$. At the same time, {asking that the effective optical depth $\tau_{eff}=[(\tau_{abs}+\tau_{es})\cdot \tau_{abs}]^{1/2} < 1$ ($\tau_{abs}$ is the true emission-absorption optical depth) and assuming $\tau_{abs}<<\tau_{es}$, we obtain $r_{cor} \ga 15$ cm} (again for $kT_{cor}\approx 1$ keV, $\tau_{es} \approx 5$). Therefore, physical conditions are consistent with the existence of both compact and moderately extended optically thick coronae.

{We note that the sources occupy both the \textit{very thick} and \textit{thick} regions of the $\tau-kT_{cor}$ plane (divided by the boundary at $\tau \sim 9$, as nominally defined in \citealt{pintore12}). However, the two regions do not appear to be completely detached on this plane but rather smoothly connected. Some sources are also residing in either one or the other state at different times (see below).

Concerning the soft component, the temperatures are consistent with those found in previous works ($\sim 0.15-0.30$ keV, e.g. \citealt{stobbart06}). Only in some observations of NGC 253 X-1, we found that temperature of the soft component increases up to 0.8 keV, although it shows equally good fits with both a cold and a warm disc component \citep{barnard10}}.
We emphasize that the soft component may contribute $\sim 20-30\%$ of the total emission in the 0.3-10 keV energy range for the less luminous sources and up to $\sim 50\%$ for the most luminous ones. This suggests either a higher obscuration of the hard component or an increase in the importance of the soft component, both likely associated to the onset of stronger winds.

\subsection{Chemical abundances}
\begin{table}
  \begin{center}
    \scalebox{0.85}{\begin{minipage}{15.0cm}
   \caption{Abundances inferred from the Oxygen K-shell photoionization edge (0.538 keV).}
   \label{chem_abund}
   \begin{tabular}{l c c c c}
\hline 
&\multicolumn{4}{c}{\textbf{$12+\log(O/H)^a$}}\\
\\
No. Observation &  $\#2$ & $\#3$& $\#4$& \\ 
\\
IC 342 X-1 & $8.75\pm0.13$ & $8.75\pm0.15$ & $8.63\pm0.15$ &\\  
\hline
No. Observation& $\#2$&&  & \\
\\
NGC 253 X-1 & $8.8\pm0.1$ &  &  &\\
\hline
No. Observation& $\#7$& $\#8$ & $\#9$& $\#10$\\
\\
NGC 5408 X-1 & $8.71_{-0.11}^{+0.7}$ & $8.72_{-0.16}^{+0.10}$ & $8.64_{-0.08}^{+0.12}$ & $8.63_{-0.04}^{+0.17}$\\
\hline
No. Observation& $\#3$& $\#7$&  & \\
\\
Ho IX X-1 & $8.73_{-0.04}^{+0.05}$ & $8.68_{-0.16}^{+0.11}$ &  &\\
\hline
No. Observation& $\#5$&&  & \\
\\
Ho II X-1 & $8.63_{-0.06}^{+0.14}$ &  &  &\\
\hline
\end{tabular}
\end{minipage}}
\end{center}
$^a$ For the solar abundance we assume $12+\log(O/H)=8.69$ \citep{asplund09}.
\end{table}

\begin{table*}
  \begin{center}
   \caption{Fractional variability of the ULXs analysed in this work.}
    \scalebox{0.85}{\begin{minipage}{14.0cm}
\footnotesize
   \label{timing}
   \begin{tabular}{l c c c c c c c }
\hline 
 \hline
Source & Obs. ID & \multicolumn{2}{c}{$0.3-10$ keV} & \multicolumn{2}{c}{$0.3-2.0$ keV} &\multicolumn{2}{c}{$2.0-10$ keV} \\
\\
& & counts s$^{-1}$ & $F_{var}^a$ & counts s$^{-1}$ & $F_{var}^a$& counts s$^{-1}$ & $F_{var}^a$  \\
\hline
\\
\multirow{4}{*}{IC 342 X-1} & 0093640901  &$0.490\pm 0.010$ & $\leq13$& $0.248\pm 0.009$ & $\leq19$&$0.252\pm 0.009$ & $2\pm19$\\
 & 0206890101  & $1.090\pm 0.010$ & $7\pm 2$& $0.548\pm 0.009$ & $\leq13$& $0.550\pm 0.009$ & $8 \pm 3$ \\
 & 0206890201&  $0.561\pm 0.007$ & $6\pm 2$ & $0.294\pm 0.005$ & $2\pm8$&$0.276\pm 0.005$ & $8 \pm 3$\\
 & 0206890401  &  $1.430\pm 0.020$ & $21 \pm 2$& $0.67\pm 0.01$ & $12\pm 3$& $0.77\pm 0.02$ & $29\pm 2$\\
\hline
\\
\multirow{7}{*}{NGC 253 X-1} & 0110900101  & $0.146\pm 0.007$ & $\leq40$& $0.132\pm 0.006$ & $\leq39$& $0.066\pm 0.007$ & $ \leq95$\\
 & 0152020101 &  $0.408\pm 0.004$ & $29\pm 1$& $0.416\pm 0.004$ & $25 \pm 1$& $0.137\pm 0.003$ & $43\pm 3$\\
 & 0304851101&  $0.148\pm 0.006$ & $\leq31$ &  $0.121\pm 0.006$ & $\leq40$&   $0.070\pm 0.005$ & $\leq68$\\
 & 0304850901 &  $0.180\pm 0.006$ & $12\pm5$&  $0.137\pm 0.005$ & $\leq23$&  $0.077\pm 0.005$ & $\leq47$\\
 & 0304851001 &  $0.199\pm 0.006$ & $\leq18$&  $0.152\pm 0.005$ & $\leq23$&  $0.082\pm 0.005$ & $\leq45$\\
 & 0304851201 & $0.205\pm 0.005$ & $\leq18$& $0.157\pm 0.004$ & $\leq20$&  $0.077\pm 0.003$ & $\leq41$\\
 & 0304851301&    $0.209\pm 0.009$ & $\leq27$&  $0.157\pm 0.008$ & $\leq29$ &  $0.085\pm 0.007$ & $\leq52$\\
\hline
\\
\multirow{7}{*}{Ho IX X-1} & 0112521001 & $2.150\pm 0.020$ & $1\pm 3$& $1.55\pm 0.02$ & $4\pm2$& $0.62\pm0.01$ & $\leq12$\\
 & 0112521101 &  $2.500\pm 0.020$ & $\leq5$& $1.78\pm 0.02$ & $\leq6$& $0.72\pm 0.01$ & $\leq10$\\
 & 0200980101 &  $1.709\pm 0.006$ & $2\pm1$&  $1.185\pm 0.005$ & $\leq5$&  $0.537\pm 0.004$ & $2\pm3$\\
 & 0657801601 & $5.000\pm 0.100$ & $1\pm21$& $2.6\pm 0.1$ & $\leq21$& $1.48\pm0.07$ & $\leq21$\\
 & 0657801801 &  $3.680\pm 0.050$ & $1\pm7$& $2.56\pm 0.04$ & $\leq12$& $1.16\pm 0.03$ & $23\pm3$\\
 & 0657802001 &  $2.370\pm 0.040$ & $\leq9$&  $1.68\pm 0.03$ & $4\pm3$&  $0.74\pm 0.02$ & $\leq15$\\
 & 0657802201 & $4.770\pm 0.030$ & $\leq6$& $3.25\pm 0.03$ & $\leq7$& $1.56\pm0.02$ & $\leq10$\\
\hline
\\
\multirow{6}{*}{NGC 5204 X-1} & 0142770101 & $0.667\pm 0.008$ & $5\pm 2$ & $0.557\pm 0.007$ & $5 \pm 2$& $0.121\pm0.003$ & $\leq24$\\
 & 0142770301 &  $0.920\pm 0.020$ & $\leq13$& $0.79\pm 0.02$ & $\leq14$& $0.146\pm 0.008$ & $\leq37$\\
 & 0150650301 &  $1.090\pm 0.020$ & $\leq12$&  $0.95\pm 0.02$ & $\leq14$&  $0.147\pm 0.008$ & $\leq34$\\
 & 0405690101& $1.370\pm 0.020$ & $\leq8$ & $1.20\pm 0.02$ & $\leq9$ & $0.176\pm0.007$& $\leq26$\\
 & 0405690201&  $1.118\pm 0.007$ & $4\pm1$ & $0.975\pm 0.006$ & $5\pm1$& $0.154\pm 0.003$ & $\leq14$\\
 & 0405690501 &  $0.839\pm 0.007$ & $\leq7$&  $0.714\pm 0.006$ & $\leq7$&  $0.135\pm 0.003$ & $\leq19$\\
\hline
\\
\multirow{8}{*}{NGC 5408 X-1} & 0112290501& $1.440\pm 0.030$ & $17\pm 2$& $1.37\pm 0.02$ & $16 \pm 2$ & $0.094\pm0.007$ & $\leq46$\\
 & 0112290601 &  $1.410\pm 0.020$ & $5\pm 2$& $1.34\pm 0.02$ & $5\pm2$ & $0.091\pm 0.006$ & $\leq38$\\
 & 0112291201 &$-$ & $-$&$-$ & $-$&$-$ & $-$\\
 & 0302900101 & $1.070\pm 0.004$ & $8.8\pm 0.5$& $0.995\pm 0.004$ & $8.1 \pm 0.5$& $0.094\pm0.002$ & $\leq24$\\
 & 0500750101 &  $1.000\pm 0.006$ & $12.9\pm0.7$& $0.924\pm 0.006$ & $11.3\pm0.8$& $0.102\pm 0.002$ & $\leq25$\\
 & 0653380201&  $1.200\pm 0.004$ & $7.0\pm 0.5$ &  $1.110\pm 0.004$ & $6.9 \pm 0.5$& $0.111\pm0.002$ & $\leq20$\\
 & 0653380301 & $1.182\pm 0.004$ & $6.8\pm 0.4$& $1.091\pm 0.004$ & $6.8\pm0.5$& $0.109\pm 0.001$ & $\leq18$\\
 & 0653380401 & $1.123\pm 0.004$ & $7.7\pm 0.5$& $1.038\pm 0.004$ & $6.9\pm 0.5$& $0.105\pm 0.002$ & $\leq21$\\
 & 0653380501 &  $1.084\pm 0.004$ & $9.0\pm 0.5$&  $0.997\pm 0.004$ & $7.9\pm 0.5$&  $0.107\pm 0.002$ & $\leq19$\\
\hline
\\
\multirow{5}{*}{Ho II X-1} & 0112520601& $3.370\pm 0.030$ & $3\pm 1$ & $2.97\pm 0.03$ & $4 \pm 1$& $0.40\pm0.01$ & $\leq14$\\
 & 0112520701 &  $3.100\pm 0.100$ & $\leq21$& $2.6\pm 0.1$ & $\leq25$&  $0.43\pm 0.02$ & $\leq22$\\
 & 0112520901 &  $0.880\pm 0.020$ & $2\pm6$& $0.81\pm 0.02$ & $2\pm7$&  $0.090\pm 0.006$ & $\leq40$\\
 & 0200470101 &  $3.390\pm 0.010$ & $2.6\pm 0.9$& $2.98\pm 0.01$ & $3.8\pm 0.8$&  $0.423\pm 0.005$ & $\leq13$\\
 & 0561580401 &  $1.330\pm 0.010$ & $21.4\pm 0.7$& $1.201\pm 0.009$ & $21.5\pm 0.8$&  $0.143\pm 0.003$ & $\leq18$\\
\hline
\end{tabular}
\end{minipage}}
\end{center}
$^a$ Calculated from the background subtracted EPIC-pn light curves, with 200 s time bins. Values without error bars indicate 3$\sigma$ upper limits.\\
\end{table*}

We tentatively tried to determine the chemical abundances in the ULX environments looking for  Oxygen and Iron edges in their X-ray spectra. In the spectral fits the \textit{tbabs} absorption model is replaced with \textit{tbvarabs} that allows variation in the chemical abundances (and grain composition). We set alternatively the abundance of Oxygen or Iron to zero. The spectrum was then fitted with the EPIC continuum best fitting model (keeping all parameters, but normalizations, fixed) plus an absorption edge, that accounts for the observed absorption feature. {For NGC 5408 X-1, we removed the APEC component which may significantly affect the measurement}. The parameters of the edge are then used to compute the abundance (assuming for the solar Oxygen metallicity the value 8.69 dex; \citealt{asplund09}). We selected only EPIC-pn spectra with more than 5000 counts \citep{winter07}, i.e. observations $\#2,3,4$ for IC 342 X-1, $\#2$ for NGC 253 X-1, $\#$3, 7 for Holmberg IX X-1, $\#$1,4,5,6 for NGC 5204 X-1, $\#7,8,9$ and 10 for NGC 5408 X-1 and $\#5$ for Holmberg II X-1. For NGC 5204 X-1, the X-ray spectrum is rather insensitive to variations of the O abundance {so that no definite conclusion can be drawn}. 
In Table~\ref{chem_abund} we show the estimated chemical abundances and their uncertainties. 
{The abundances are consistent with solar with the exception of NGC 253 X-1 which is marginally super-solar (8.8 dex). The metallicity of Ho II X-1 obtained by \citet{goad06} differs from ours possibly because of the different {reference value adopted for [O/H]$_\odot$ (8.93 dex, \citealt{wilms00}).}}

Finally, no clear evidence of an {iron L-shell edge was found in any of the sources of our sample. This is in line with the lack of evidence of the iron K-shell features in ULX spectra coming from an ionised wind (e.g. \citealt{walton13}).}

\subsection{Temporal analysis}
\label{temporal_analysis}

We complemented the spectral analysis with an investigation of the temporal properties of the ULXs of our sample. For each source we computed the fractional root mean square (RMS) variability amplitude ($F_{var}$), that measures the variance of a source over the Poissonian noise in the time domain and is usually normalized to the average count-rate (\citealt{edelson02,vaughan03}). The fractional variability can be used also to study the energy dependence of the short-term variability of the source and it is a powerful tool in order to characterise the properties of the spectral components (e.g. \citealt{middleton11}a). We evaluated $F_{var}$ from background subtracted lightcurves in several energy bands $-$ 0.3-10 keV, 0.3-2.0 keV and 2.0-10 keV $-$ binning them in intervals of $\Delta T=200$s. This is a good compromise for all the sources because it allows to have at least 20 counts in each time bin and at least 20 bins.
In Table~\ref{timing} we report the fractional variability measurements for all the observations. When the statistics are not sufficient we provide only the $3\sigma$ upper limit.

{Unlike \citet{sutton13} who selected only the highest quality observations for each source of their sample, we study the variability of all the observations. We find that the 0.3-10 keV band fractional is between a few percent and $\sim 30\%$, with marginal evidence for higher values in the harder band. However, the variability is not clearly correlated with any specific spectral regime, varying almost randomly across different observations and from source to source. \citet{sutton13} found a possible correlation between $F_{var}$ and the spectral shape, showing that a higher variability is seen during the \textit{soft ultraluminous} state, {i.e. when the soft component is dominant}. They suggested that such short term variability is related to turbulences at the edge of the outflow ejected by the disc and would then be larger when our line of sight intersects the edge of the wind. This would imply that high levels of variability can be seen only in favourable epochs when the wind opening angle crosses our line of sight. According to \citet{sutton13}, the occurrence of this condition is more likely in the \textit{soft ultraluminous} state, when the wind is more extended and its opening angle smaller. However, among the three sources of our sample with a stronger soft component, only NGC 5408 X-1 shows very high levels of short-term variability. In the other two cases the variability is less than 10$\%$. Therefore, we do not find significant evidence for this effect, indicating that in \textit{soft ultraluminous} state the wind opening angle might assume a larger range of values.}

Finally we mention that in the last observation of IC 342 X-1 the variability is definitely larger than the average $F_{var}$ ($\sim10\%$), and stronger at higher energies ($\sim$10$\%$ at $0.3-2.0$ keV against $\sim30\%$ at $2.0-10.0$ keV; Table~\ref{timing}). This is probably caused by a significant drop in flux occurring during the observation. A similar behaviour was observed also in the last observation (\#5) of Holmberg II X-1, during which the source switches from a high to a low flux level with a decrement of a factor of 2 (\citealt{kajava12}). On the other hand, NGC 253 X-1 shows also higher than average $F_{var}$ but in this case the variability is intrinsic and, notably, not related to any significant flux change. This finding makes this source different from the other ULXs of our sample (see next section).

\subsection{Looking at single sources}

The spectral analysis reported above shows that the spectra of ULXs can be well described in terms of a disc plus comptonisation model in which the {Comptonising} component is usually optically thick and cool. The sources occupy preferentially two different regions of the $\tau-kT_{cor}$ plane (\textit{thick} and \textit{very thick state}). 
{IC 342 X-1 and NGC 253 X-1 populate typically the \textit{very thick} state, while Holmberg II X-1, Holmberg IX X-1, NGC 5204 X-1 and NGC 5408 X-1 stay predominantly in the \textit{thick} state. Only NGC 1313 X-1 and X-2 appear to cross convincingly both regions, while Holmberg IX X-1 and NGC 5204 X-1 may do it only marginally.}

As mentioned above, NGC 5408 X-1 and Holmberg II X-1 show prominent soft components in their spectra. In NGC 5408 X-1 the significance of the soft component is slightly dependent on the total luminosity. At variance with the general trend, the soft component contributes almost 40-50$\%$ of the total flux ($L_{disc}\sim (3-4)\cdot10^{39}$ erg s$^{-1}$) when the source is in a low luminosity state and becomes less significant ($\sim 25\%$ of the total flux) when the luminosity increases ($L_{disc}\sim 6-7\cdot 10^{39}$ erg s$^{-1}$). We also found evidence for a disc luminosity-temperature relation of the type $L_{disc} \propto T_{disc}^{1.8\pm 0.8}$. Short term variability is generally present at low energies but essentially unconstrained at high energies because of the low statistics\footnote{Different values of the fractional RMS of NGC 5408 X-1 were found by \citealt{caballero13}, probably because of the different choice of the time sampling.}.
The presence of a strong soft component and significant short-term variability in NGC 5408 X-1 may be explained within the scenario in which the wind component becomes very extended, its opening angle narrower and its edge  intersects our line of sight, consistent with what proposed by \citet{middleton11}b and \citet{sutton13}.

In Holmberg II X-1, we found a disc luminosity-temperature relation of the type $L_{disc}\propto T_{disc}^{3.6\pm1.4}$. While the correlation of Holmberg II X-1 may be reminiscent of a standard disc relation, {we do not interpret it in this way because in the} model adopted here the corona is optically thick and the temperature of the soft component refers either to the outer visible part of the disc or to the wind.

\subsubsection{NGC 253 X-1}
\label{subsec_ngc253}
{The behaviour of NGC 253 X-1 appears slightly different from the other ULXs of our sample}. It shows a significantly curved spectrum and its short term variability may reach $\sim$20-30$\%$. As shown in Table~\ref{timing}, the RMS fractional variability in the highest quality observation (which is also the more luminous) is $\sim 30\%$ in the $0.3-10$ keV energy band. The counting statistics of observations $\#2$ and 4 allows us to study the RMS spectrum of the source, that was calculated by selecting background subtracted EPIC-pn lightcurves in the energy ranges 0.3-0.5, 0.5-0.7, 0.7-1.0, 1.0-1.3, 1.3-1.6, 1.6-2.1, 2.1-4.0, 4.0-10 keV (Figure~\ref{figrms253}). {When fitted with a constant the spectrum of the fractional variability is consistent with a value of $\sim 28\%$.} These findings suggest that the observed emission may come from a single component. {However, we note that, although not statistically significant, in observation $\#2$ there may be some hint of variability: $F_{var}$ is $\sim 25\%$ in the 0.3-2.0 keV band and $\sim 40\%$ in the 2.0-10 keV band (see also \citealt{sutton13}). This fact may be consistent with the existence of an additional component affecting the high energy part of the spectrum.
{If we interpret it as an absorption component, it may be produced by material that originates from the impact of the accretion stream onto the disc itself or by a low-density, equatorial wind ({compare with} \citealt{sutton13}). This material may be responsible also for the short decrements of the count rate observed by \citet{barnard10} (Figure 2 in his paper), that may appear similar to the dips observed in some XRBs \citep[e.g.][]{white82,diaz06}. In this scenario the source would be seen at rather large inclinations. 

\begin{figure}
\center
\subfigure{\includegraphics[height=4.5cm,width=6.4cm]{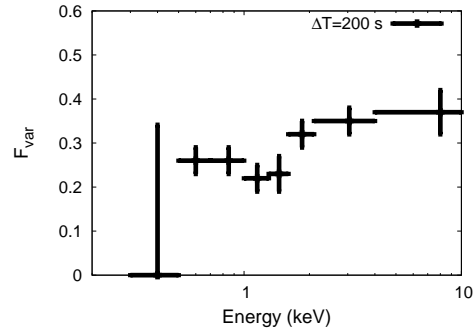}}
\caption{NGC 253 X-1: RMS fractional variability spectrum evaluated on background subtracted EPIC-pn lightcurves of observation 0152020101, sampled with time bins of 200 s. A fit with a constant value ($F_{var}\sim 28\%$) is consistent with the spectrum.} 
\label{figrms253}
\end{figure}

\begin{figure}
\center
\includegraphics[height=5.4cm,width=7.7cm]{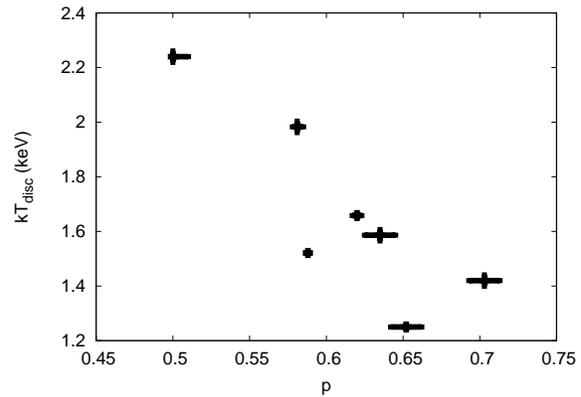}
\caption{
NGC 253 X-1: inner disc temperature versus \textit{p}-index for the \textit{diskpbb} model fit.}
\label{fig_conf_IC342}
\end{figure}

\begin{table}
\footnotesize
   \caption{Best fitting spectral parameters of NGC 253 X-1 obtained with the absorbed {slim disc model (\textit{diskpbb} in Xspec)}. Error bars are at $90\%$ for each parameter of interest.}
\begin{center}
\scalebox{0.78}{\begin{minipage}{14.0cm}
   \label{ccd_ngc253_slim}
   \begin{tabular}{l c c c c c}
\hline
\multicolumn{6}{c}{ NGC 253 X-1}  \\ 
\hline
No. & $N_H$$^a$ &  $p$$^b$  &$kT_{disc}$$^{c}$ & $L_{X}$ [0.3-10 keV] $^d$ & $\chi^2/dof$ \\ 
& ($10^{21}$ cm$^2$) & (keV) &  & ($10^{38}$ erg s$^{-1}$) & \\
\hline
1& $0.9_{-0.01}^{+0.01}$	& $0.5_{-0.5}^{+0.001}$ & $2.24_{-0.02}^{+0.02}$ & $8.3_{-0.8}^{+0.6}$ & 117.81/99\\
2& $1.46_{-0.04}^{+0.04}$     &  $0.588_{-0.001}^{+0.001}$ & $1.521_{-0.004}^{+0.004}$ & $27_{-1}^{+1}$ & 856.21/812\\  
3& $0.4_{-0.1}^{+0.1}$	&  $0.652_{-0.01}^{+0.01}$ & $1.25_{-0.01}^{+0.01}$ & $7.0_{-0.6}^{+0.7}$ & 90/93\\
4& $\leq0.2$	&  $0.703_{-0.01}^{+0.01}$ & $1.42_{-0.02}^{+0.02}$ & $10_{-1}^{+1}$ & 75.88/75\\ 
5& $1.2_{-0.2}^{+0.2}$ & $ 0.581_{-0.003}^{+0.003}$ & $1.98_{-0.02}^{+0.02}$ & $13_{-9}^{+2}$ & 88.97/85\\
6& $1.0_{-0.1}^{+0.1}$	&  $0.620_{-0.003}^{+0.003}$ & $1.66_{-0.01}^{+0.01}$ & $13.0_{-1}^{+1}$ & 174.94/163\\
7& $0.9_{-0.2}^{+0.2}$ & $0.635_{-0.01}^{+0.01}$ & $1.59_{-0.02}^{+0.02}$ & $13.0_{-2}^{+2}$ & 44.38/47\\ 
\hline
   \end{tabular} 
\end{minipage}}
 \end{center}
$^a$ Column density; $^b$ exponent of the radial dependence of the disc temperature; $^c$ disc temperature; $^d$ unabsorbed total X-ray luminosity in the 0.3 -10 keV range; 
\end{table}

NGC 253 X-1 differs from the other sources of our sample also because its X-ray spectrum can be well described by a { modified/slim} disc model. 
Indeed, this source was {classified as \textit{broadened disc} by \citet{sutton13}}. The {modified/slim} disc model incorporates the effects that are expected to set in at or slightly above the Eddington limit and was modelled through a \textit{diskpbb} component\footnote{This model depends on the inner disc temperature $kT_{disc}$ and a parameter \textit{p} that describes the radial profile of the disc temperature as $T_{disc}\propto r^{-p}$, where $p$=0.75 for a standard disc and $p$=0.5 for a slim disc.}. Although the fits are generally statistically acceptable (Table~\ref{ccd_ngc253_slim}), we found one observations (\#4) in which the column density pegged at 0 indicating that the absorption was anomalously low. However, fixing the column density at the average value of the other observations, the \textit{diskpbb} model provides still a good description of the spectrum ($\chi^2/dof=80.61/76$, $kT_{disc}\sim1.76$ and $p\sim0.6$).
The variation of the inner disc temperature $kT_{disc}$ with the parameter \textit{p} of the \textit{diskpbb} model is shown in Figure~\ref{fig_conf_IC342}. There is a tendency for the $p$-index to decrease with increasing disc temperature, so that the higher is the temperature the more advection dominated is the disc. 
At low luminosity (observation $\#1,3$), the spectrum can also be well described in terms of a canonical \textit{diskbb+powerlaw} (with $kT_{disc}\sim 0.3$ keV and $\Gamma \sim 2.2$, $\chi^2/dof=92.80/98$) or \textit{diskbb} spectral model \citep[with $kT_{disc}=1.1$ keV, $\chi^2/dof=92.29/94$; see also][]{kajava09}. 
Hence this source, which is also the faintest of our sample, may be in a state similar to the \textit{soft/steep powerlaw} state of Galactic XRBs and may switch to the ULX regime only at high luminosity.

\section{colours}
\label{colorrr}

\begin{figure*}
\center
\subfigure{\includegraphics[height=9cm,width=12cm]{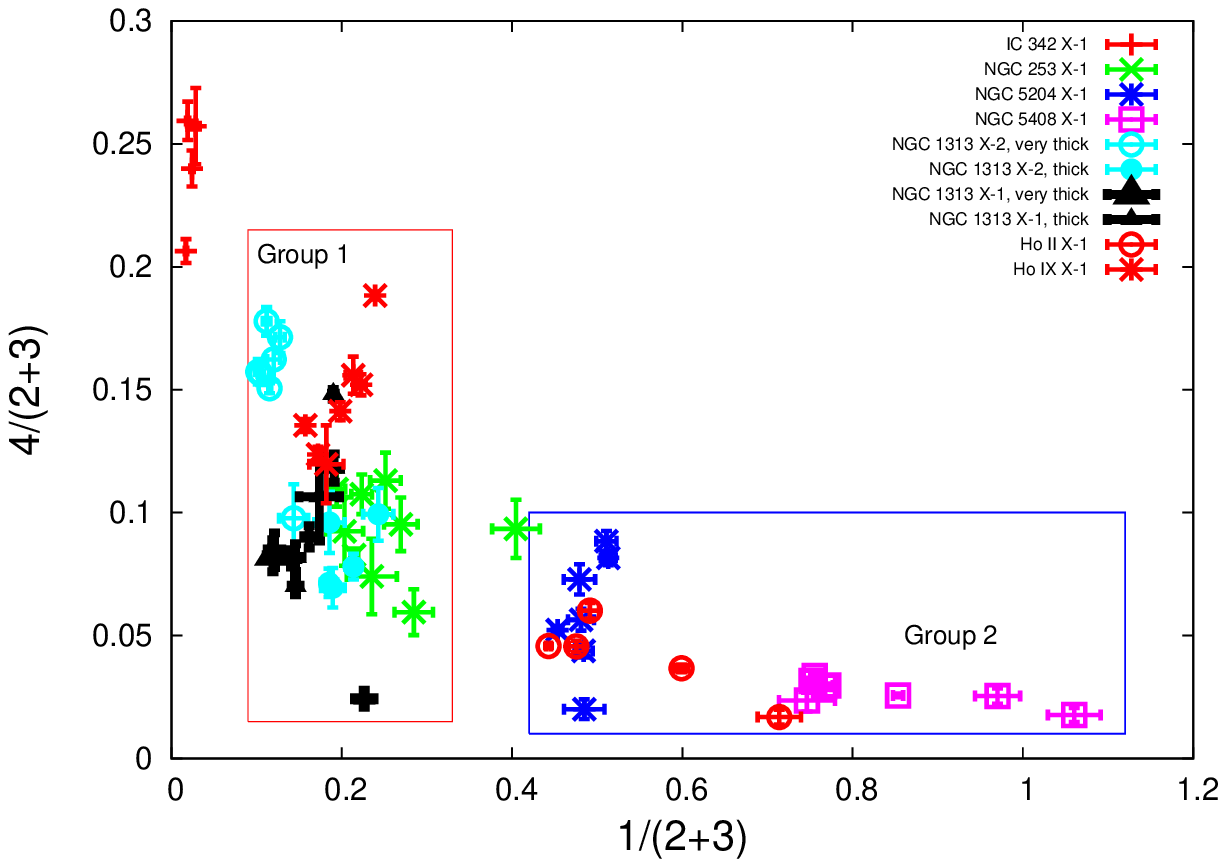}}
\subfigure{\includegraphics[height=9cm,width=12cm]{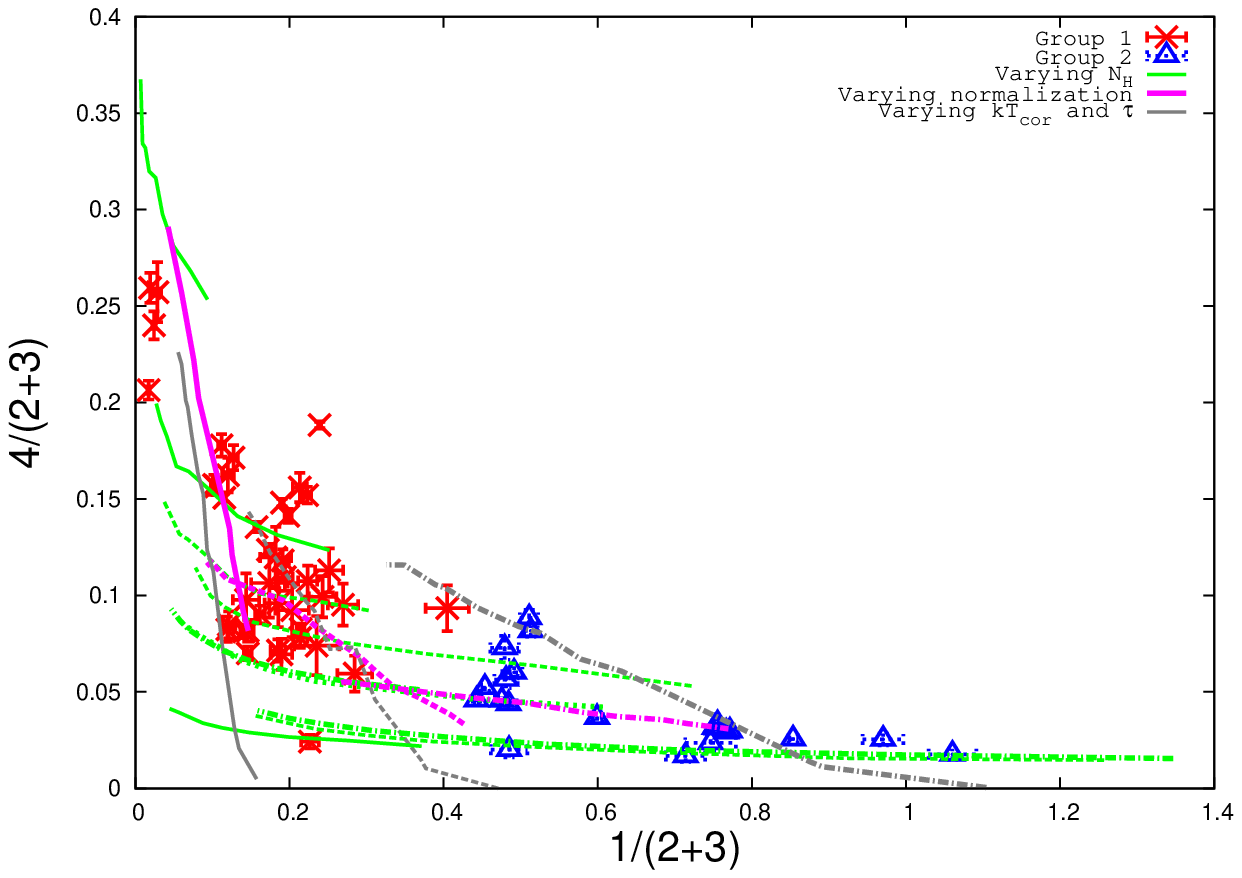}}
\caption{Colour-colour (\textit{top}) diagram obtained using the hardness ratios defined by the energy bands 1 {(0.3-0.7 keV), 2 (0.7-2.0 keV), 3 (2.0-4.0 keV) and 4 (4.0-10 keV)} (see text). {The sources appear to cluster in two different regions. {We grouped them according to the value of the 1/(2+3) colour: 0.1-0.3 for group 1 (\textit{red} box) and 0.4-1.1 for group 2 (\textit{blue} box).} (\textit{bottom}) The same colour-colour diagrams with the tracks computed from \textit{diskbb+comptt} spectra with varying column density, normalization of the soft component and parameters of the hard component (as explained in the text). Sources of group 1 are marked with red crosses, while those of group 2 with blue triangles. The \textit{green} lines represent tracks with varying $N_H$ (decreasing from left to right). Lines with the same hatch refer to different (fixed) values of the normalization of the soft component (see text). The \textit{solid/dotted/dashed} lines are computed starting from the best fit parameters of two models of group 1 (very-thick and thick state) and one of group 2, respectively.
The \textit{magenta} lines are computed starting from the same spectral fits, varying the normalization of the soft component (that increases from left to right). Finally, the three \textit{grey} lines are computed varying the parameters of the comptonising corona. Observations in group 1 are mostly related by a combined variation of $N_H$ and the parameters of the high energy component. Instead, the transition from group 1 to group 2 is mainly driven by variations of the normalization of the soft component, consistent with either an increment of the soft component or a decrement of absorption.}} 
\label{fig_color_sensible}
\end{figure*}

In the previous Section we have shown that almost all the spectra of the ULXs of our sample can be described by a combination of a soft component and a comptonisation model. The spectral parameters span different ranges but, if interpreted at face value, indicate similar physical conditions, such as an optically thick {cold} corona and a {cool} disc. 
{However, the lowest counting statistics spectra do not provide strong constraints on the spectral fits because of the degeneracy of some spectral parameters, especially if the temperature of the disc and that of the soft photon input are free to vary independently.}
In fact, this situation is rather common for most of the available \textit{XMM-Newton} observations of ULXs and {also for simpler models, as the \textit{diskbb+powerlaw} for example}. 
{Hence we tried to support the findings of the spectral analysis using a complementary approach based on the \textit{hardness ratios}}. This technique is very powerful for low counting statistics data and appears then particularly suitable for many observations of ULXs. In this section we will {reconsider} all the observations using this approach, that in the future may be adopted also for analysing larger samples of lower counting statistics data.

The method of the \textit{hardness ratios} or \textit{colour diagrams} has been successfully adopted in the past to study the behaviour of XRBs and, more in general, of X-ray sources \citep{maccacaro88}. Extensive monitoring of some Galactic BH binaries with \textit{Rossi-XTE} led to the discovery of a common evolutionary path in the hardness-intensity diagram, the hysteresis cycle, that all the BH binary systems accreting at sub-Eddington rates appear to follow (see for example \citealt{belloni10}). This cycle describes how the X-ray spectrum changes with the source {intensity}.

A \textit{hardness} ratio can be defined as $B_i/B_j$, in which $B_i$ and $B_j$ are the total counts in two given energy bands.
Since different instruments have different response matrices, because of the higher throughput we used only EPIC-pn data and selected {four energy bands: 0.3-0.7, 0.7-2.0, 2.0-4.0 and 4.0-10 keV, defined for simplicity as $1, 2, 3$ and $4$, respectively.} 
This choice provides an adequate sampling of the spectral regions in which the soft excess, spectral pivoting (see \citealt{kajava09,pintore12}) and/or the comptonizing component are usually contributing. In addition, it guarantees a similar counting statistics in every band {and it allows us to discriminate in a clearer way the differences in the spectral properties of our sources (see below)}. In each energy range of interest, we added the counts of every good channel of the spectrum and subtracted the total counts of the corresponding background channels, properly scaled for the extraction areas.
{Count rates are evaluated dividing the counts by the net EPIC-pn GTI per observation and are then rescaled to a distance of 1 Mpc (errors on the distances are not taken into account).}
{We add the NGC 1313 X-1 and X-2 data \citep{pintore12} to the sample for comparison, finding a similar spectral evolution.}

{Figure~\ref{fig_color_sensible}-\textit{top} shows a \textit{colour-colour} diagram in which the $y$-axis is defined as the ratio between the counts in the energy bands 4 and (2+3) and the $x$-axis as the ratio between the bands 1 and (2+3)}. The sources lie along a sequence starting from IC 342 X-1 and ending with NGC 5408 X-1. Apart from IC 342 X-1 and letting aside the details of the evolution of single sources, {we note that the scatter is not very large. This may suggest that differences induced by the likely diverse BH masses and inclination of the sources are not so strong and that the BH masses themselves are not very different. }
At least \textit{two} groups of observations can be identified {on the colour-colour plot}: group 1 {with $1/(2+3)\sim 0.1-0.3$ and group 2 with $1/(2+3)\sim 0.4-1.1$}. In group 1, there are the less luminous and most absorbed sources: NGC 1313 X-1 and X-2, Ho IX X-1 and NGC 253 X-1 (although one of its observations is also consistent with belonging to group 2). The observations of NGC 1313 X-2 are clearly split in two subgroups which are reminiscent of the \textit{very thick} and \textit{thick} states identified in previous works \citep{feng09,pintore12}. {A similar trend can also be observed for NGC 1313 X-1 and Holmberg IX X-1.} 

{Group 2 contains the most luminous and less absorbed sources $-$ NGC 5204 X-1, Ho II X-1 and NGC 5408 X-1 $-$ where a strong soft component is seen. Using a different spectral selection and analysis, \citet{sutton13} classified ULXs in three spectral states/groups of sources that they called \textit{broadened disc, hard ultraluminous} and \textit{soft ultraluminous}. {We suggest that the \textit{broadened disc} and \textit{hard ultraluminous} sources can be identified with group 1 and cannot be easily distinguished as their X-ray colours are similar. On the other hand the \textit{soft ultraluminous} state can be associated to group 2. \textit{Therefore, the spectral groups found via a detailed spectral analysis can be identified equally well adopting a model-independent colour-based approach, making use of colour-colour diagrams incredibly powerful also for poor quality data.}}}
In group 2, NGC 5204 X-1 shows variations in the 4/(2+3) colour up to a factor of 3 while 1/(2+3) stays nearly constant ($\sim0.5$), indicating either that the high energy component is significantly variable or that the soft bands vary together.
On the other hand, the $1/(2+3)$ colour of NGC 5408 X-1 changes {by almost a factor of 2}, while 4/(2+3) remains nearly constant, meaning that variability is mostly in the soft component or {that the harder energy bands vary together.}

The only source that appears clearly separated from the others in the colour-colour diagram is IC 342 X-1. We know that, in terms of spectral parameters, IC 342 X-1 shows similarities with the \textit{very thick} state of NGC 1313 X-2. On the other hand, IC 342 X-1 has the largest intrinsic  $N_{H}$ of the whole sample and hence its position on the colour-colour diagram is strongly affected by absorption at soft energies which modifies the hardness ratios.

\subsection{Spectral evolution and colours}
\begin{figure*}
\center
\subfigure{\includegraphics[height=6.9cm,width=8.8cm]{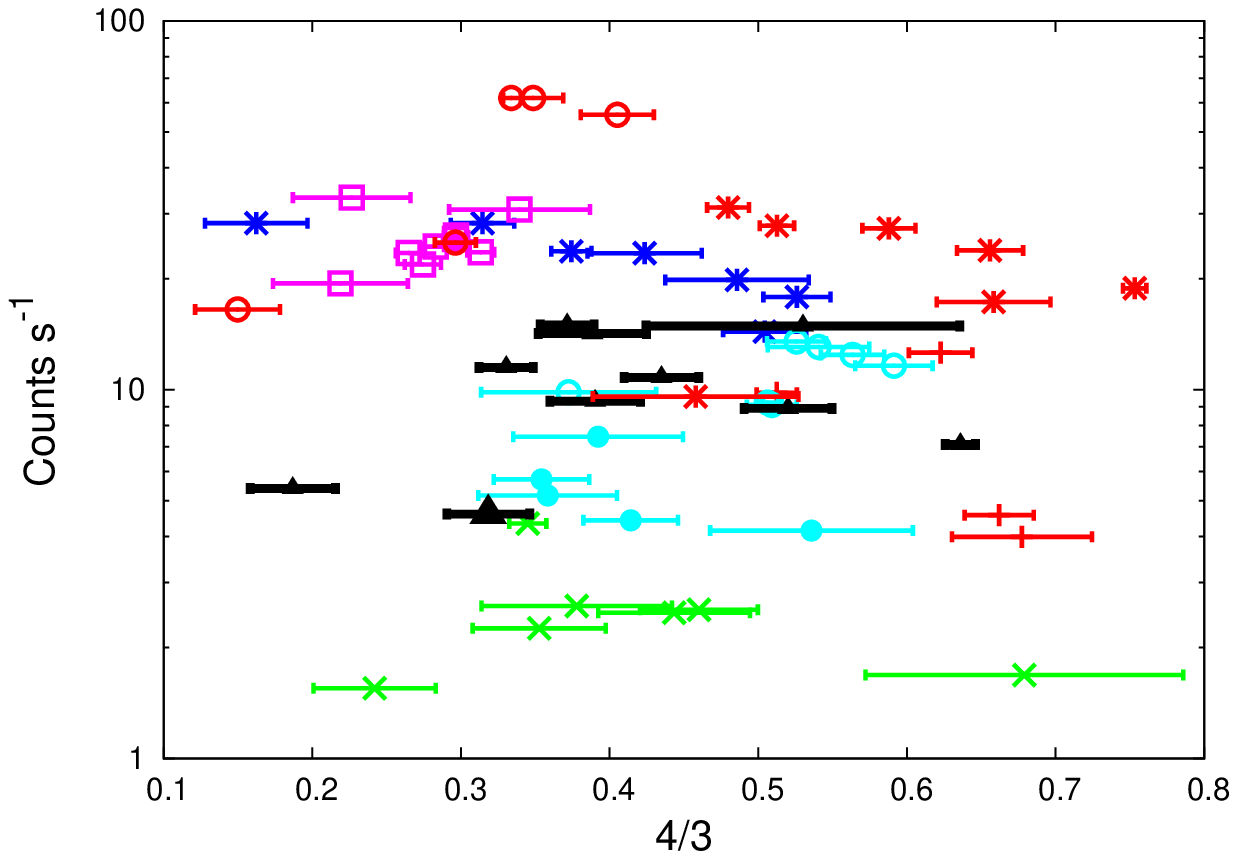}}
\subfigure{\includegraphics[height=6.9cm,width=8.8cm]{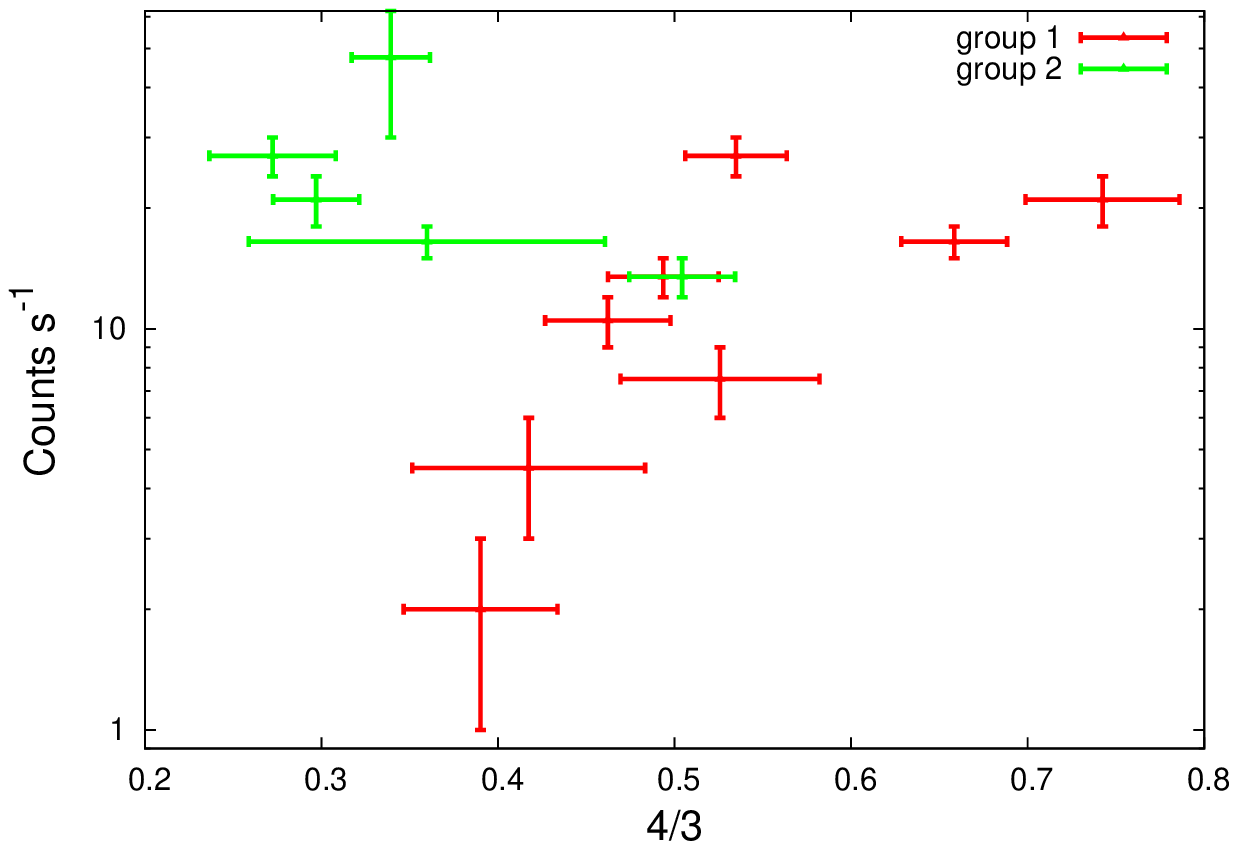}}
\caption{(\textit{left}) HID of all the EPIC-pn data of our sample of ULXs. The count rate is evaluated in the 0.3-10 keV energy range and rescaled to a distance of 1 Mpc. Count rates vary up to a factor of $\sim 100$. Energy bands 3 and 4 refer to 2.0-4.0 and 4.0-10 keV, respectively. The color coding refers to the legenda in Figure~\ref{fig_color_sensible}. (\textit{right}) HID rebinned in terms of count rate for group 1 (\textit{red}) and group 2 (\textit{green}). To reduce the scatter in the data, bins of 3 counts s$^{-1}$ and 6 counts s$^{-1}$ were adopted below and above 15 counts s$^{-1}$, respectively.}
\label{hardness_rebin}
\end{figure*}

{In order to understand if the positions of the sources of the two groups on the colour-colour diagram are related, we tried to explore the possible paths of the spectral evolution on this plane and to understand what is the main component which drives the variability.}
We selected the parameters of three different models, the first two fitted to observations of group 1 and one fitted to an observation of group 2.
Starting from the best fitting parameters of these observations (that are reported in Table~\ref{ic342_table}), we changed {step by step each single parameter and simulated a fake spectrum. We then computed the colours of these fake spectra and reported them on the colour-colour plot, joining the various points with lines}. The tracks computed in this way are shown in Figure~\ref{fig_color_sensible}-\textit{bottom}. We vary the values of the intrinsic column density, the normalization of the soft component and the parameters of the hard component (keeping fixed the Compton parameter $y=\tau^2 4kT_{cor}/m_e c^2\sim0.75$). {For $N_H$ we sample the range $10^{20}-6.2\cdot10^{21}$ cm$^{-2}$ with steps of $8\cdot10^{20}$ cm$^{-2}$.
For the normalization of the soft component we consider values from 1\% to 600$\%$ of the best fit, with steps of 0.1-60 (XSPEC units).
For the parameters of the corona we span the range $kT_{cor}=0.5-6$ and $\tau=14-4$, with incremental steps of $0.4-2$ keV for the electron temperature and $0.5-4$ for $\tau$ (at the top of each track kT$_{cor}=$0.5 and $\tau=14$).
The tracks for varying $N_H$ are computed fixing the normalization of the soft component at three different values (best fit, 10 times smaller and 10 times larger).
}
{We note that, with the definition of the colours adopted here, a spectral model with fixed $N_H$ and a single varying parameter is represented by a straight line on the colour-colour plot}. 

The main parameters driving the observed changes in the position of the sources of group 1 are likely to be the normalization of the soft component and the parameters of the corona ({electron temperature and optical depth}; see magenta and gray tracks, Figure~\ref{fig_color_sensible}-\textit{bottom}). Also variations in $N_H$ (green tracks, Figure~\ref{fig_color_sensible}-\textit{bottom}) or changes in more than a single parameters may account for the variability patterns observed in group 1. {As discussed in the next Section, we interpret the changes in the parameters of the corona in terms of variations of density/temperature induced by variations in the accretion rate. Different levels of absorption do also play a role and are related to the extension of the wind that starts to develop and to the viewing angle (e.g. IC 342 X-1; see below)}.

{On the other hand, the transition from group \textit{1} to group \textit{2} can be explained mostly by variations in $N_{H}$ and/or in the normalization of the soft component, while changes in the parameters of the corona appears less consistent with the observed behaviour (see again Figure~\ref{fig_color_sensible}-\textit{bottom}).
This is consistent with the scenario proposed by \citet{gladstone09} and \citet{sutton13} in which the soft component becomes more and more significant along a certain spectral sequence which here appears to correspond to the path from group 1 to group 2. This transition can be explained in terms of the increasing importance of wind emission as we enter in group 2 and is in
agreement with the results of hydrodynamical simulations of super-Eddington accretion (e.g. \citealt{ohsuga09}).}

The tracks in Figure~\ref{fig_color_sensible}-\textit{bottom} account also for the isolated position of IC 342 X-1. The region where it resides is connected to sources of group 1 through lines of decreasing $N_H$. To test this, we simulated spectra with the best fitting \textit{diskbb+comptt} parameters of the longest observation of NGC 1313 X-2 (\citealt{pintore12}) and changed the value of $N_{H}$ setting it equal to $6\cdot10^{21}$ cm$^{-2}$ (consistent with the mean value of IC 342 X-1) and leaving the other parameters unchanged. Then we evaluated its \textit{colours}: they fall on the location occupied by the observations of IC 342 X-1, i.e. $4/(2+3)\sim0.25$ and $1/(2+3)\sim0.05$. Therefore, the two sources are in a similar spectral state, the main difference being related to the different level of absorption of the soft component, the latter associated to the photosphere of the wind.

\subsection{Hardness-intensity diagrams}
\label{HID}
{In the following, we use the same energy bands introduced in Section 4, i.e. 0.3-1.0, 1.0-2.0, 2.0-4.0 and 4.0-10 keV, labeled as 1, 2, 3 and 4. With the choice of the energy bands adopted in this work}, the hardness ratio $4/3$ is essentially unaffected by variations of $N_{H}$. We use this hardness ratio to construct a hardness-intensity diagram, plotted in Figure~\ref{hardness_rebin}-\textit{left}.
The total absorbed count rates in the 0.3-10 keV energy band {are evaluated adding the counts in each band and dividing them by the net EPIC-pn GTI} per observation. These rates are then rescaled to a distance of 1 Mpc {(uncertainties on the distances are not taken into account)}. 

We suggest that, all together, the spectra of ULXs may define a path on the hardness-intensity plane (Figure~\ref{hardness_rebin}, \textit{left}): low luminosity sources (as NGC 253 X-1 and NGC 1313 X-2) show a slight tendency to harden as the count rate increases up to the position of the \textit{very thick} state of NGC 1313 X-2, IC 342 X-1 and some very thick observations of Ho IX X-1. This is caused by the development of a bell-shaped spectrum with a flattening below 5 keV and a pronounced curvature above this energy (see \citealt{pintore12}). At higher count rates, Ho IX X-1 and NGC 5204 X-1 show an opposite behaviour, becoming softer when their flux increases. At the extreme of this sequence there are Ho II X-1 and NGC 5408 X-1 that tend to show even softer ${4/3}$ colours as the count rate slightly decreases.
To better clarify this pattern, we rebinned separately the data of the sources in group 1 and 2 in bins of 3 counts s$^{-1}$ below 15 counts s$^{-1}$ and 6 counts s$^{-1}$ above 15 counts s$^{-1}$. This turns out to be the best compromise between effectively reducing the scatter in hardness ratios and tracking the ULX colour evolution with sufficient accuracy}. In each bin, we averaged the colours and count rates. 
The resulting plot is shown in Figure~\ref{hardness_rebin}-\textit{right}.
The two groups mentioned previously and the spectral path described above may be better identified on this diagram.

\section{Discussion}
\label{discussion}

We have shown that, although a common reference model can be used to describe the spectral properties of ULXs, the sources show different spectral evolutions. These changes can be phenomenologically described in terms of variations in the properties of a soft component and a high energy tail. Variations at low energies are accounted for (mostly) by changes in the normalization of a soft component and/or in the column density to the source, while variations in the high energy tail by changes in the parameters of an optically thick corona. 
This spectral variability is rather well characterized on a hardness-intensity diagram in terms of a suitably defined hardness ratio (Figure \ref{hardness_rebin}). {We have shown the existence of two groups of sources characterised by different properties of the high energy component and diverse colours in the soft bands (group 1 and group 2). 
Recently, \citet{sutton13} identified three spectral groups in ULXs and classified in terms of their spectral shapes (\textit{broadened disc, hard ultraluminous} and \textit{soft ultraluminous}). They computed hardness-intensity diagrams using the flux estimated from the spectral analysis.
We suggest that the \textit{broadened disc} and \textit{hard ultraluminous} sources can be identified with group 1, while the \textit{soft ultraluminous} state can be associated to group 2. 
Using the approach presented here sources can be more easily classified in terms of the 1/(2+3) colour, although the actual boundary between the two states (1/(2+3)$\sim 0.3-0.4$) may vary in future when considering a larger sample of sources.}

We are then led to qualitatively speculate on the physical origin of the observed spectral phenomenology of the sources of our sample in light of recent work done in this area. Assuming that the black hole masses in the different sources are comparable, the main physical parameter that drives the variability along the path in Figure \ref{fig_color_sensible} may be the accretion rate, increasing (in Eddington units) from NGC 253 X-1 (bottom) to {Ho II X-1} (top-left). In particular we suggest that there is a turning point after which the spectral evolution and the accretion mechanism change character. From Figure \ref{hardness_rebin}, we estimate that the 0.3-10 keV count rate of the turning point is {$\sim 10-20$ EPIC-pn counts s$^{-1}$}, at a distance of 1 Mpc. {Assuming an absorbed powerlaw spectrum with Galactic/intrinsic $N_H$ of $3.9\cdot 10^{20}$ cm$^{-2}$/ $10^{21}$ cm$^{-2}$ and $\Gamma=2.2$ (average of the spectral index for the observations around this point), this corresponds to $L_X\sim 4-8\cdot10^{39}$ erg s$^{-1}$. Setting it equal to the Eddington luminosity would correspond to a mass in the range $\sim 30-60 \, M_\odot$. However, as discussed below, we cannot use this value to straightforwardly estimate the BH mass.}

\begin{figure*}
\center
\includegraphics[height=5cm,width=10cm]{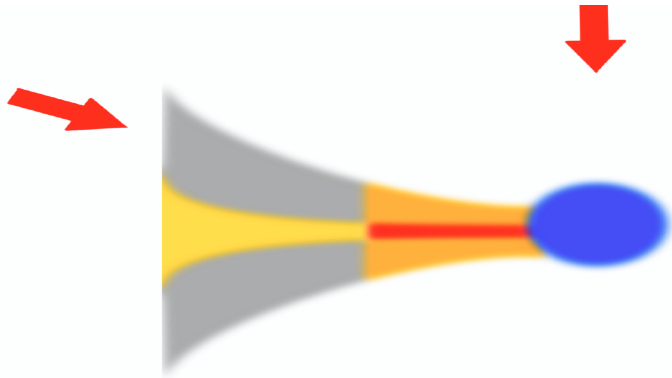}
\caption{A schematic representation of the geometry of the inflow/outflow in a ULX. The yellow/red area is the accretion disc, the \textit{grey} region is the optically thin phase of the wind while the orange area represents the region in which the wind is ejected and produces the soft emission. Finally, the blue area is the innermost region of the disc possibly coupled to an optically thick corona or the comptonised disc itself. The red arrows represent a face-on and a high inclination angle of the line of sight.
} 
\label{fig_accretion}
\end{figure*}

{In agreement with the results of previous spectral analyses \citep[e.g.][]{kajava09}}, {we find that, even adopting the \textit{diskbb+comptt} model}, the soft component does not show the characteristic correlation between temperature and luminosity expected for a standard accretion disc. {This further suggests} that this component is more likely associated to strong and extended outflows which may set in at the high accretion rates expected in ULXs (see e.g. \citealt{ohsuga09,gladstone09,kajava09}). The wind may be sufficiently dense and extended to cover a large fraction of the outer regions of the disc and be in part responsible for the different absorption levels we observe in the sources of our sample. If the inclination angle to the source is large (edge-on=90$^{\circ}$, face-on=0$^{\circ}$), our line of sight passes through the {outer}, colder regions of the outflow (see e.g. \citealt{middleton11}b; \citealt{sutton13}) and we observe a higher column density {that significantly absorbs} the soft component\footnote{We remind the reader that the absorption component to which we refer here is intrinsic, as the Galactic absorption in the direction of the sources is separately taken into account in the spectral fits.}. The accretion disc {is then} buried under the wind or an optically thick comptonizing medium in the innermost regions from which the high energy spectral component may originate (see e.g. \citealt{middleton11b}a,b). {The spectrum could also be modified by self-Comptonisation in the inner radiation-pressure dominated region}. A schematic representation of the geometry of this inflow/outflow is shown in Figure~\ref{fig_accretion}. A somewhat similar representation for the origin of equatorial winds in the soft state of BH XRBs has been proposed by \citet{ponti12}. However, in ULXs the extreme accretion rate may cause ejections of the wind also from the inner regions of the disc and the onset of turbulence. The wind in ULXs can be also more mass loaded and hence become optically thick (\citealt{poutanen07,ohsuga09,middleton11b}b).

Variations in the accretion rate may induce variations in the density/size of the wind and temperature/optical depth of the innermost {regions}, leading to the observed X-ray spectral variability.
{In this context}, sources of group 1 could represent the low-density wind tail of the distribution. {In these sources,} the outflow is present but not copious enough to remove a significant amount of energy from the inner regions, which appear {progressively different} when the accretion rate and the total flux increase. 
We can try to explain the evolutionary path from the \textit{thick} to the \textit{very thick} spectral state in group 1 within this scenario. As the accretion rate increases, the wind and the corona become progressively more mass loaded and part of the disc internal energy goes into the wind, reducing the temperature of the corona {and increasing its optical depth}. 
{The only source in group 1 that may be interpreted in a different way is NGC 253 X-1, as its spectral fits do not need a soft/wind component.
As discussed in section~\ref{subsec_ngc253}, we think that the spectra of this ULX are better described by an advection dominated disc. A further hint that this source is in a different regime comes also from the fact that its position on the colour-colour diagram is significantly displaced from the \textit{very thick} state of NGC 1313 X-2. }

On the other hand, when the accretion rate increases further, radiative forces become strong enough to blow out a larger amount of material from the disc. This can represent the physical regime of the sources in group 2. {Our spectral analysis suggests that in these ULXs the soft/wind component can reach 50$\%$ of the total flux and is, in general, coupled to a high energy component with less pronounced curvature than sources in group 1. The larger luminosity of the sources in group 2 could be due in part to the larger accretion rate and in part to the soft emission arising from the optically thick phase of the wind, that makes the soft component more pronounced (e.g. \citealt{gladstone09}). In this regime a greater amount of material is removed from the innermost regions of the accretion flow and the corona, making its optical depth lower and reducing the energy lost by Compton cooling on the electrons. The corona/hard component appears then hotter than in group 1. {This causes a change of the spectral shape from curved to powerlaw-like. As the optical depth decreases, the spectra become steeper, explaining the inversion of the trend of the hardness ratio with the total count rate observed in the hardness-intensity diagram (Figure \ref{hardness_rebin}).}} At the same time, a fraction of radiative energy in the disc is converted in kinetic energy to power the outflow. {Evidence for the existence of a compact and optically thick corona has also been recently found in another bright ($> 10^{40}$ erg s$^{-1}$) ULX (Circinus ULX5) with \textit{NuSTAR} (\citealt{walton13})}.

The transition from the regime of group 1 to that of group 2 corresponds to the observed turning point in the hardness-intensity diagram at a count rate of $\sim 10-20$ count s$^{-1}$. NGC 1313 X-1 and X-2, NGC 5204 X-1 and possibly Ho IX X-1 seem to switch between the two accretion regimes. 

In this scenario intrinsic absorption and inclination are related. IC 342 X-1 is the most absorbed source and hence our line of sight could be fully intersecting the denser part of the wind.
In NGC 1313 X-1 and X-2 the inclination could be relatively smaller, but our line of sight still intersecting the neutral phase of the wind causing some absorption. 
On the other hand, sources in group 2 show on average a smaller $N_{H}$ and hence we probably see them at smaller inclinations or {their winds are predominantly ionised}. At present it is unclear if this caused by the small statistics of our sample or by selection effects. {In the latter case, sources with more extended, ionised winds may have the characteristic spectra of group 2 because, at small inclinations, the inner hard emission is still visible and not behind the wind photosphere.}

As mentioned above, in some observations of a few sources we find short term variability at the level of $\leq 10$\%, with no clear dependence on the energy band, but in other cases the variability is well constrained at significantly lower levels. \citet{middleton11b}b and \citet{sutton13} proposed that high values of variability at high energy may be due to the clumpy nature of the wind along our line of sight if it intersects its outer edge. High levels of variability would be seen only when the line of sight encounters wandering blobs of optically thick matter in the wind. 
{Our results suggests that the short-term variability is not a specific characteristic of one of the spectral groups but, instead, it appears as a stochastic effect which can be observed when our line of sight crosses the turbulences at the edge of the wind.}

Therefore, a combination of variations in accretion rate, inclination angle/absorption and mass of the BH appears able to qualitatively account for the key properties of the X-ray spectral evolution of the present sample of ULXs within a framework of a non-conventional Eddington/super-Eddington accretion regime. High counting statistics observations of other ULXs and a more extended analysis of the {colour-colour and} hardness-intensity diagram for a larger sample of ULXs will be able to confirm or disprove the interpretation proposed above.

\section*{Acknowledgements} 
We would like to thank the anonymous referee for his/her constructive criticisms and useful comments and suggestions. 
We also would like to thank Tim Roberts for his advice on dealing with the spectral and temporal analysis, and Gabriele Ponti for useful discussions on the search for correlations between the temporal and spectral properties. We acknowledge financial support through INAF grant PRIN-2011-1 (Challenging Ultraluminous X-ray sources: chasing their black holes and formation pathways) and INAF grant PRIN 2012-6.

\addcontentsline{toc}{section}{Bibliography}
\bibliographystyle{mn2e}
\bibliography{biblio}

\end{document}